\documentclass[12pt]{article}
\usepackage{latexsym,epsf,cite}

\makeatletter
\@addtoreset{equation}{section}

\makeatother

\begin{document} 

\setlength{\unitlength}{0.2cm}

\title{
End-to-end distribution function for dilute polymers
}
\author{
  \\
  {\small Sergio Caracciolo}             \\[-0.2cm]
  {\small\it 
           Scuola Normale Superiore and INFN -- Sezione di Pisa}  \\[-0.2cm]
  {\small\it I-56100 Pisa, ITALY}         \\[-0.2cm]
  {\small Internet: {\tt Sergio.Caracciolo@sns.it}}     \\[-0.2cm]
        \\[-0.1cm]  \and
  {\small Maria Serena Causo}             \\[-0.2cm]
  {\small\it John von Neumann-Institut f\"ur Computing (NIC)}  \\[-0.2cm]
  {\small\it Forschungszentrum J\"ulich}        \\[-0.2cm]
  {\small\it D-52425 J\"ulich, GERMANY}          \\[-0.2cm]
  {\small Internet: {\tt M.S.Causo@fz-juelich.de}}     \\[-0.2cm]
  \\[-0.1cm]  \and
  {\small Andrea Pelissetto}                          \\[-0.2cm]
  {\small\it Dipartimento di Fisica and INFN -- Sezione di Roma I} \\[-0.2cm]
  {\small\it Universit\`a degli Studi di Roma ``La Sapienza"}        \\[-0.2cm]
  {\small\it I-00185 Roma, ITALY}          \\[-0.2cm]
  {\small Internet: {\tt Andrea.Pelissetto@roma1.infn.it}}   \\[-0.2cm]
  {\protect\makebox[5in]{\quad}}  
  \\
}
\vspace{0.5cm}

\maketitle
\thispagestyle{empty}   

\vspace{0.2cm}

\begin{abstract}
We study the end-to-end distribution function for dilute polymers. 
We present a computation to order $O(\epsilon^2)$, $\epsilon = 4 - d$, 
and discuss in detail its asymptotic behaviour for small 
and large distances. The theoretical predictions are compared with 
Monte Carlo results, finding good agreement.
We show that the McKenzie-Moore-des Cloizeaux
phenomelogical ansatz provides a very precise approximation to the exact EEDF.
\end{abstract}

\clearpage

\newcommand{\be}{\begin{equation}}
\newcommand{\ee}{\end{equation}}
\newcommand{\beq}{\begin{equation}}
\newcommand{\eeq}{\end{equation}}
\newcommand{\bea}{\begin{eqnarray}}
\newcommand{\eea}{\end{eqnarray}}
\newcommand{\<}{\langle}
\renewcommand{\>}{\rangle}

\def\spose#1{\hbox to 0pt{#1\hss}}
\def\ltapprox{\mathrel{\spose{\lower 3pt\hbox{$\mathchar"218$}}
 \raise 2.0pt\hbox{$\mathchar"13C$}}}
\def\gtapprox{\mathrel{\spose{\lower 3pt\hbox{$\mathchar"218$}}
 \raise 2.0pt\hbox{$\mathchar"13E$}}}
\def\lle{<<}
\def\gge{>>}

\newcommand{\R}{\hbox{{\rm I}\kern-.2em\hbox{\rm R}}}
\def\brho{\mbox{\protect\boldmath $\rho$}}

\newcommand{\reff}[1]{(\ref{#1})}
\def\smfrac#1#2{{\textstyle\frac{#1}{#2}}}

\section{Introduction}

The statistical properties of dilute polymers in good solvents have been
the subject of extensive studies during the years 
\cite{Flory_book,DeGennes_book,DesCloizeaux-Jannink_book,Freed_book}. 
A significant understanding
of the problem was reached when it was realized that long polymers 
could be modelled by chains with an excluded-volume
interaction. This allowed the introduction of simplified
theoretical models which could be analyzed more easily.
{}From a theoretical point of view, an important step forward was
made by de Gennes \cite{DeGennes_72}, 
who proved that the statistical properties 
of polymers could be obtained as the limit $N\to 0$ of the 
$N$-component $\phi^4$ theory, opening the field to the many
methods that were developed at the time  for the study 
of the critical behaviour of spin systems.

In this paper we consider the end-to-end distribution function (EEDF).
During the last thirty years a lot of work has been devoted to 
the study of this quantity. Exact results were obtained 
in Refs. \cite{Fisher-Hiley_61,Fisher_66,Mazur_65,McKenzie-Moore_71,%
McKenzie_76,DesCloizeaux_74_80}, and many numerical studies 
checked some of  the theoretical predictions
\cite{Verdier-Stockmayer_62,Wall-etal_63,Baumgartner_81,%
Bishop-Clarke_91,Bishop-etal_91,%
Dayantis-Palierne_91,Eizenberg-Klafter_93,Pedersen-etal_96,Valleau_96}.
In particular, these works tried to understand which of the 
several phenomenological expressions \cite{footnote0} provided
the best description of the numerical data. 
The proposal by Mazur \cite{Mazur_65} was clearly excluded 
\cite{Bishop-Clarke_91}, while the theoretically motivated form 
of McKenzie and Moore \cite{McKenzie-Moore_71} and
des Cloizeaux \cite{DesCloizeaux_74_80,DesCloizeaux-Jannink_book}
was confirmed to a good accuracy
\cite{Dayantis-Palierne_91,Eizenberg-Klafter_93,Valleau_96}.
It is interesting to notice that a precise 
knowledge of the EEDF could be relevant in experimental studies.
Indeed, as observed by des Cloizeaux and Jannink 
\cite{DesCloizeaux-Jannink_book},
the EEDF could be determined from scattering experiments with 
a dilute solution of polymers with marked endpoints. A 
measurement of the scattered intensity at large angles would determine
the EEDF in the large-momentum region. This would provide an estimate of the 
critical exponent $\gamma$, which is otherwise inaccessible 
experimentally. 

In this paper we reconsider the problem of the determination of the 
EEDF. We extend the $\epsilon$-expansion calculations of 
Ref. \cite{Oono-etal_81} to order $\epsilon^2$ and give perturbative
expansions for the quantities that characterize 
the asymptotic behaviour for small and 
large distances. Using the $\epsilon$-expansion we can show that 
the phenomenological parametrization 
of \cite{McKenzie-Moore_71,DesCloizeaux_74_80} is essentially exact for
distances much larger than the correlation length
--- the expected discrepancy is of order 1-5\% in the region accessible 
to simulations ---
while in the opposite range the discrepancy should be (at most) of order
10\%. We also give theoretical expressions for several quantities
derived from the EEDF. Using precise Monte Carlo estimates of the 
critical exponents, we derive accurate theoretical predictions 
for the asymptotic behaviour of the EEDF.  The theoretical estimates are 
compared with numerical results obtained from
a simulation of self-avoiding walks on a cubic lattice,
finding good agreement.

The paper is organized as follows:
in Sec. 2 we introduce our notations and definitions and review the exact 
results that are available for the EEDF. In Sec. 3 we report our
computation of the EEDF to order $\epsilon^2$. Only the results are 
given, the technical details being presented in the Appendix.
In Sec. 4 we obtain estimates for the asymptotic behaviour of the 
EEDF using the Laplace-de Gennes transform. Finally in Sec. 5
we discuss the numerical results.

\section{Definitions} \label{sec2}

We consider a monodisperse ensemble of polymers with $N$ monomers. 
If ${\bf r}$ is the vector joining the endpoints of the walk, 
we will be interested in computing the
unnormalized distribution $c_N({\bf r})$ \cite{footnote1}
of the endpoint vector. We also introduce 
a normalized distribution
\be
P_N({\bf r}) = {c_N({\bf r})\over \sum_r c_N({\bf r})}\; ,
\ee
the mean squared end-to-end distance 
\be
    R^2_{e,N} = \sum_r |{\bf r}|^2 P_N({\bf r}),
\ee
and the related correlation length
\be
\xi_N^2 = {1\over 2d} R^2_{e,N}.
\label{eq2.3}
\ee
In the limit $N\to\infty$, $|{\bf r}|\to \infty$, 
with $|{\bf r}| N^{-\nu}\sim |{\bf r}|/\xi_N$ fixed, 
the function $P_N({\bf r})$ has the
scaling form \cite{Fisher_66,McKenzie-Moore_71,DesCloizeaux_74_80}
\be
P_N({\bf r}) \approx {1\over \xi^d_{N}} f(\rho) 
   \left[1 + O(N^{-\Delta})\right],
\label{deffrho}
\ee
where $\brho = {\bf r}/\xi_{N}$, $\rho = |\brho|$, $d$ is the space
dimensionality, and $\Delta$ is a correction-to-scaling exponent.
By definition
\bea
   \int_0^\infty \rho^{d-1} d\rho\, f(\rho) &=& {1\over S_d}\; ,
\label{norm1} \\
   \int_0^\infty \rho^{d+1} d\rho\, f(\rho) &=& {2 d\over S_d}\; ,
\label{norm2}
\eea
where $S_d$ is the volume of the $d$-dimensional sphere
\be
S_d = {2 \pi^{d/2}\over \Gamma(d/2)} \;\; .
\ee
Several facts are known about $f(\rho)$. For large values of $\rho$ 
it behaves as 
\cite{Fisher-Hiley_61,Fisher_66,McKenzie-Moore_71,DesCloizeaux_74_80}
\be
f(\rho) \,\approx f_\infty \rho^\sigma
   \exp\left(-D \rho^\delta\right)\; ,
\label{flargerho}
\ee
where $\sigma$ and $\delta$ are given by
\bea
   \delta &=& {1\over 1-\nu}, \label{delta} \\
   \sigma &=& {2 \nu d - 2 \gamma + 2 - d \over 2 (1-\nu)}.
\eea
For $\rho\to 0$, we have \cite{McKenzie-Moore_71,DesCloizeaux_74_80}
\be
   f(\rho) \approx f_0 \left({\rho\over2}\right)^\theta,
\label{fsmallrho}
\ee
where
\be
\theta =\, {\gamma - 1\over \nu}\; .  \label{theta} 
\ee
We can also consider the Fourier transform of $f(\rho)$,
\be
\widetilde{f}(Q) = \, \int {d^d {\brho}\over (2\pi)^d} \,
   e^{i {\bf Q}\cdot {\brho}} \, f(\rho),
\ee
which is the critical (large-$N$) limit of 
$\widetilde{c}_N({\bf q})/\widetilde{c}_N({\bf 0})$ 
with ${\bf Q} \equiv {\bf q}\xi_N$ fixed, $\widetilde{c}_N({\bf q})$ 
being the Fourier transform of $c_N({\bf r})$. 
For $Q\to 0$, $\widetilde{f}(Q)$ has a regular expansion in powers 
of $Q^2$, while for $Q^2\to \infty$ it behaves as 
\be
\widetilde{f}(Q) = \widetilde{f}_\infty Q^{-\theta - d}.
\label{ftildelargeQ}
\ee
The constants $\widetilde{f}_\infty$ and $f_0$
are related. Indeed
\be
f_0 =\, \widetilde{f}_\infty\, (4 \pi)^{-d/2}
{\Gamma\left(-{\theta\over2}\right)\over 
 \Gamma\left({\theta+d\over2}\right)}.
\label{relf0ftildeinf}
\ee
A phenomenological representation for the function $f(\rho)$
has been proposed by McKenzie and Moore
\cite{McKenzie-Moore_71} 
and des Cloizeaux \cite{DesCloizeaux-Jannink_book}:
\be
f(\rho) \approx {f}_{\rm ph}(\rho) =
   f_{\rm ph} \rho^\theta
    \exp\left(-{D}_{\rm ph} \rho^\delta\right).
\label{frhoapprox}
\ee
Here $\delta$ and $\theta$ are fixed by \reff{delta} and \reff{theta},
while ${f}_{\rm ph}$ and ${D}_{\rm ph}$ 
are fixed by the normalization conditions
\reff{norm1} and \reff{norm2}:
\bea
D_{\rm ph} &=& \left\{ {\Gamma[(1-\nu)(2 + d +\theta)] \over 2 d\,
                      \Gamma[(1-\nu)(d +\theta)] } \right\}^{\delta\over2}
    \; ,\label{Dph}\\
f_{\rm ph} &=& {\delta {D}_{\rm ph}^{(d +\theta)(1-\nu)} \over S_d\,
                   \Gamma[(1-\nu)(d +\theta)]} \; .\label{fph}
\eea
In two dimensions
\bea
D_{\rm ph} &=& 0.026339478 \ldots ,\\
f_{\rm ph} &=& 0.046757638 \ldots ,
\eea
while in three dimensions, using the precise estimate
$\nu = 0.58758 \pm 0.00007$ \cite{Belohorec-Nickel_97} and
our result \cite{Caracciolo-etal_97} $\gamma = 1.1575 \pm 0.0006$, we have
\bea
D_{\rm ph} &=& 0.14470  \pm 0.00014 , \label{Dph-d3}\\
f_{\rm ph} &=& 0.015990 \pm 0.000008 , \label{fph-d3}\\
\theta &=& 0.2680 \pm 0.0011,\\
\delta &=& 2.4247 \pm 0.0004 .
\eea
Notice that $f_{\rm ph} (\rho)$ cannot be exact since the exponents
$\theta$ and $\sigma$ are different, but this is less crucial
in $d=3$, since their numerical values are very similar.
In order to improve the approximation, one can also take $\theta$ and 
$\delta$ as free parameters, and determine them by fitting 
numerical estimates of $f(\rho)$. However, once $\theta$ and $\delta$
are given, $D_{\rm ph}$ and $f_{\rm ph}$ are uniquely determined by
\reff{Dph} and \reff{fph}. 
In the following we will indicate with ``phenomenological representation"
the function \reff{frhoapprox} with $\delta$ and $\theta$ fixed to 
their theoretical values.

For the purpose of computing $D$ and $\delta$ from Monte Carlo data,
it is much easier to consider the ``wall-to-wall"
distribution function
\be
P_{N,w}(x) = \sum_{x_2,\ldots,x_d} P_N(x,x_2,\ldots,x_d),
\ee
which represents the probability that the endpoint of the walk lies on
a plane at a distance $x$ from the
origin of the walk. In the large-$N$ limit, $P_{N,w}(x)$ has the scaling form
\be
P_{N,w}(x) = {1\over \xi_{N}} f_w(\rho) \left(1 + O(N^{-\Delta})\right), 
\ee
where $\rho = x/\xi_N$.
We will show in Sec. \ref{sec4} that for large $\rho$ we have
\be
f_w(\rho) \approx f_{w,\infty} \rho^{\sigma_w}
     \exp(- D \rho^\delta)\; ,
\label{Eq2.27}
\ee
where $\delta$ is given by \reff{delta}, $D$ is the same constant appearing in 
Eq. \reff{flargerho}, and
\be
 \sigma_w = {\delta}\left(\nu - \gamma + {1\over2}\right).
\label{Eq2.28}
\ee

\section{Distribution function to order $O(\epsilon^2)$} \label{sec3}

We will now derive the EEDF $f(\rho)$
using a continuum description and the standard $\epsilon$-expansion.
We start from the Edwards' path integral \cite{Edwards_65}
for the probability distribution
function of the end-to-end distance ${\bf r}$ of a chain with contour
length $N$ in $d$ space dimensions
\be
c_N({\bf r}) =\, 
  \int_{{\bf x}(0) = {\bf 0}}^{{\bf x}(N) = {\bf r}} 
  D[{\bf x}]\exp\left
  \{- {1\over4} \int_0^N ds \left({d{\bf x}(s)\over ds}\right)^2
   - {w\over2} \int_0^N ds\,\int_0^N dt \, 
   \delta[{\bf x}(s) - {\bf x}(t)]\right\}.
\ee
Here ${\bf x}(s)$ is the position vector of the arc-length position $s$ 
of the chain, the integral over $D[{\bf x}]$ represents the summation over
all possible configurations between the two ends of the chain, and
$w$ is the unrenormalized strength of the excluded-volume interaction.
We use adimensional units setting the Kuhn step length equal to $2d$.

\begin{figure}
\begin{center}
\epsfxsize = 0.8 \textwidth
\leavevmode\epsffile{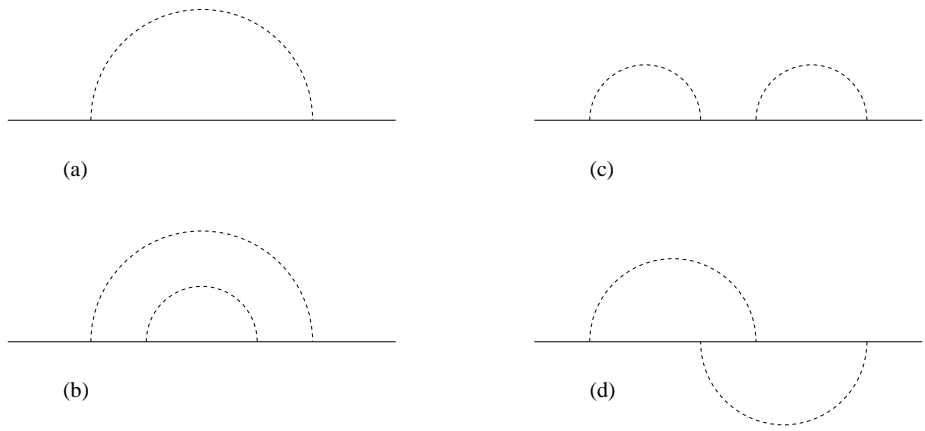}  
\end{center}
\caption{Graphs contributing to the EEDF to order $\epsilon^2$. 
 } 
\label{figuragraphs}
\end{figure}

In this Section we report the computation of the 
EEDF $f(\rho)$ to order $\epsilon^2$, where as usual
$\epsilon = 4 - d$, extending the results of 
Ref. \cite{Oono-etal_81}. The diagrams that need to 
be computed are reported in Fig. \ref{figuragraphs}. 
We obtain: 
\begin{itemize}
\item Graph (a)
\be
\hskip -1truecm
G_a(Nq^2)= 
-{1\over 2} B\left(2 - {\epsilon\over 2},-1 + {\epsilon\over 2}\right)
\left[(1 - Nq^2) e^{-Nq^2} + \epsilon g_1(Nq^2) + \epsilon^2 
   g_2(Nq^2)\right] + O(\epsilon^2);
\ee
\item Graph (b)
\bea
G_b(Nq^2) &=& {1\over 4} \left(-1 + {\epsilon\over 2}\right)
   B\left(2 - {\epsilon\over 2},-1 + {\epsilon\over 2}\right)^2 \times
\nonumber \\
&& \left[(1 - Nq^2) e^{-Nq^2} + 2 \epsilon g_1(Nq^2) + 4 \epsilon^2 
   g_2(Nq^2)\right] + O(\epsilon);
\eea
\item Graph (c)
\bea
\hskip -2truecm
G_c(Nq^2) &=& 
{1\over 4} B\left(2 - {\epsilon\over 2},-1 + {\epsilon\over 2}\right)^2
   \times
\nonumber \\
\hskip -2.8truecm
&&
  \left(1 + {N\over2} {\partial\over \partial N}\right)
 \left[(1 - Nq^2) e^{-Nq^2} + 2 \epsilon g_1(Nq^2) + 4 \epsilon^2
   g_2(Nq^2)\right] + O(\epsilon);
\eea
\item Graph (d)
\bea
\hskip -1truecm
G_d(Nq^2) &=& 
   \left[ - {3\over 2\epsilon^2} - {3\over 4\epsilon} - {3\over2} +
          {\pi^2\over 8} - {r^2\over6} -
          {3\over2} \log 2r + {3\over2} \log^2 2r\right] 
\nonumber \\
\hskip -1truecm
&& \times \left[(1 - Nq^2) e^{-Nq^2} + 
   2 \epsilon g_1(Nq^2) + 4 \epsilon^2
   g_2(Nq^2)\right] 
\nonumber \\
\hskip -1truecm
&&  - \left({1\over 8\epsilon} + {41\over96} + 
        {15\over 16 r^2} -
        {1\over4} \log 2r - {9\over 8r^2} \log 2r \right) 
        (Nq^2 e^{-Nq^2} + \epsilon g_3(Nq^2)) 
\nonumber \\ 
\hskip -1truecm
&&  \qquad +
       4 N_2(Nq^2) + 4 N_3(Nq^2) + O(\epsilon).
\eea
\end{itemize}
The functions $g_1(Nq^2)$, $g_2(Nq^2)$, $g_3(Nq^2)$, 
$N_2(Nq^2)$, and $N_3(Nq^2)$, are defined in the Appendix.
The result for diagram (d) depends apparently on 
an arbitrary parameter $r$; note that the functions 
$N_2(Nq^2)$ and $N_3(Nq^2)$ depend also on $r$ in such a way to make 
the final result independent of $r$. In principle $r$ can be set to any value.
We have kept it arbitrary, in order to have a check 
of the calculations: indeed the final results must
not depend on $r$.
At two loops we obtain for the unrenormalized $\widetilde{c}_N({\bf q})$
\be 
\widetilde{c}_N({\bf q}) = e^{-Nq^2} + \widetilde{w} N^{\epsilon/2} G_a(Nq^2) + 
    \widetilde{w}^2 N^{\epsilon} 
    \left[ G_b(Nq^2) + G_c(Nq^2) + G_d(Nq^2)\right] + 
    O(\widetilde{w}^3),
\label{cNq-perturbativo}
\ee
where $\widetilde{w} = w N_d$ and $N_d = 2 (4 \pi)^{-d/2}/\Gamma(d/2)$.
The computation of the universal EEDF goes through
several steps. First of all, we compute the correlation length 
$\xi_N$ using the definition \reff{eq2.3} and then we express $N$ in terms 
of $\xi_N$. We obtain 
\be
N =\, \xi^2_N\left [
   1 + \alpha_1 \widetilde{w}\ \xi_N^\epsilon + 
       \alpha_2 \widetilde{w}^2\ \xi_N^{2 \epsilon} + 
       O(\widetilde{w}^3)\right],
\ee
where
\bea
\alpha_1 &=& - {1\over 3\epsilon} + {1\over3} - 
     {\gamma_E\over6} + 
     {\epsilon\over 6} \left(-1 + \gamma_E - {1\over4}\gamma_E^2 - 
         {\pi^2\over24}\right),
\\
\alpha_2 &=& {5\over 18\epsilon^2} - {35\over 72\epsilon} + 
    {5\gamma_E\over 18\epsilon} - {35\over72} \gamma_E +
    {5\over 36}\gamma^2_E - {\pi^2\over54} - {5\over48 r^2} +
    {r^2\over54} 
\nonumber \\
&& + {7\over36}\log 2r + {1\over 8 r^2} \log 2r - 
    {1\over 6} \log^2 2r + {4\over9} (N_2(0) + {N'}_{\! 2}(0)).
\eea
where $\gamma_E\approx 0.5772156649$ is the Euler's constant.

Substituting this expression into Eq. \reff{cNq-perturbativo}, 
$\widetilde{c}_N({\bf q})$ becomes a series in $\widetilde{w} \xi^\epsilon_N$,
with coefficients depending on the combination
${\bf q}\xi_N\equiv {\bf Q}$. One can then compute 
$\widetilde{f}(Q) = \widetilde{c}_N({\bf q})/\widetilde{c}_N({\bf 0})$.
This quantity, once expressed in terms of $Q$, requires only a 
renormalization of the interaction strength $w$ in order to be finite. In the 
minimal subtraction scheme we have \cite{Oono,DesCloizeaux-Jannink_book}
\be
\widetilde{w} = w_R\left(1 + {4\over \epsilon} w_R + O(w_R^2)\right).
\ee
The expansion in terms of $w_R$ is finite. The critical theory is 
obtained replacing $w_R$ with its fixed-point value $w_R^*$, 
\be
w_R^* =\, {1\over4}\epsilon + {21\over128}\epsilon^2 + O(\epsilon^3).
\ee
The final result can be written in the form
\be
\widetilde{f}(Q) = \, e^{-Q^2} + \epsilon \widetilde{f}^{(1)}(Q) +
       \epsilon^2 \widetilde{f}^{(2)}(Q) + O(\epsilon^3),
\ee
where
\bea
\widetilde{f}^{(1)}(Q) &=& {1\over4} \widehat{g}_1(Q^2) ,
\\
\widetilde{f}^{(2)}(Q) &=& \left({13\over128} - {\gamma_E\over32}\right) 
   \widehat{g}_1(Q^2)
+ {1\over16} \widehat{g}_2(Q^2) - {1\over128} \widehat{g}_3(Q^2) 
+ {1\over4} \widehat{N}_2(Q^2) + {1\over4} N_3(Q^2)
\nonumber \\
&&  
+ \left({1\over16} - {\gamma_E\over32}\right) 
   Q^2 {\widehat{g}'}_{1}(Q^2) + {Q^2\over16} 
   {\widehat{g}'}_{2}(Q^2) - {\pi^2\over 768} (Q^2)^2 e^{- Q^2},
\eea
where, for any function $h(Q^2)$, we define 
\be
\widehat{h}(Q^2) = h(Q^2) - e^{-Q^2} h(0) (1 + Q^2) - e^{-Q^2} h'(0).
\label{definizione-hat}
\ee
{}From the results reported in the Appendix one can derive the asymptotic
behaviour of $\widetilde{f}(Q)$ in the limits $Q\to0$ and $Q\to\infty$. 
For $Q\to 0$ we have
\be
\widetilde{f}(Q) =\, {1\over 1 + Q^2}\left\{ 1 - (Q^2)^2 
    \left[ {1\over2} + {\epsilon\over32} + 
         \left({3\over1024} + {\pi^2\over 768} + {\widehat{b}_2\over2}\right)
     \epsilon^2 + O(\epsilon^3)\right] + O(Q^6)\right\},
\ee
where we have introduced $\widehat{b}_2$,
\be
\widehat{b}_2 = - {1\over 1536}
   \int_0^\infty dx\, x^4 K_1^3(x) \approx - 0.000235007,
\ee
where $K_1(x)$ is a Bessel function.
For $Q^2\to\infty$ we have
\be
\widetilde{f}(Q) \approx 
    {1\over (Q^2)^2}\left[
    - {\epsilon\over8} + {\epsilon^2\over 256} (11 - 12 \gamma_E) 
     - {3 \epsilon^2\over 64} \log Q^2\right] + 
    O(\epsilon^3),
\label{largeQ-expansion}
\ee
where terms of order $\log Q^2/(Q^2)^3$ have been discarded.
The function $f(\rho)$ can be derived from $\widetilde{f}(Q)$.
We will be interested in its asymptotic behaviour for $\rho\to\infty$.
Using the results of the Appendix, a lengthy computation gives
\bea
f(\rho)
   & \approx &
   {1\over 16\pi^2} + {\epsilon\over 64\pi^2} 
   \left(\log {\rho\over2} + 2 \log 4\pi  + \smfrac{1}{2} \gamma_E - 1\right)
\nonumber \\
  && + {\epsilon^2\over 512\pi^2} 
   \left[\log^2 {\rho\over2} + \left(4 \log 4\pi + \gamma_E +
    \smfrac{1}{4}\right) \log {\rho\over2} + 1 + \smfrac{19}{24}\pi^2 \right.
\nonumber \\
  && \left. + \smfrac{1}{8} \gamma_E + \smfrac{1}{4} \gamma_E^2
     - 4 \log 4\pi + 2 \gamma_E\log 4\pi + 4 \log^2 4 \pi + 4 A\right],
\label{smallR-expansion}
\eea
where terms vanishing for $\rho\to 0$ have been neglected.
The constant $A$ is defined in Eq. \reff{definizioneintA}. 
Numerically $A\approx -1.30204$.
The expressions \reff{largeQ-expansion} and \reff{smallR-expansion}
are compatible with the asymptotic behaviours \reff{ftildelargeQ} and 
\reff{fsmallrho} and give the following expansions for the constants
$\widetilde{f}_\infty$ and $f_0$:
\bea
\widetilde{f}_\infty &=& - \smfrac{1}{8}\epsilon + 
   \smfrac{1}{256} \epsilon^2 (11 - 12 \gamma_E) + O(\epsilon^3),
\\
f_0 &=& {1\over 16\pi^2} + {\epsilon\over 64\pi^2} 
    \left(2 \log 4\pi  + \smfrac{1}{2} \gamma_E - 1\right) +\,
  {\epsilon^2\over 512\pi^2}
  \left( 1 + \smfrac{19}{24}\pi^2
   + \smfrac{1}{8} \gamma_E + \smfrac{1}{4} \gamma_E^2 \right.
\nonumber \\
&& \left. \qquad\qquad
     - 4 \log 4\pi + 2 \gamma_E\log 4\pi + 4 \log^2 4 \pi + 4 A\right) +
     O(\epsilon^3).
\eea
A check of these results is provided by 
Eq. \reff{relf0ftildeinf}. It is easy to verify that the
two previous expansions satisfy Eq. \reff{relf0ftildeinf} to order
$\epsilon$. Moreover, using Eq. \reff{relf0ftildeinf}, we can 
obtain \cite{footnote2}
the contribution of order $\epsilon^3$ to $\widetilde{f}_\infty$.  
Explicitly we have
\be
\left(-\smfrac{41}{4096} + \smfrac{21}{1024} \gamma_E 
   - \smfrac{9}{1024} \gamma_E^2 - \smfrac{9}{2048}\pi^2
   + \smfrac{33}{512}\zeta(3) - \smfrac{1}{64} A\right)\epsilon^3.
\ee
Since only three terms are available, it is difficult to obtain 
reliable estimates from these expansions. 
Setting simply $\epsilon = 1$, we obtain
\be
\widetilde{f}_\infty \approx - 0.0557, \qquad \qquad
f_0 \approx 0.0176.
\label{eq3.24}
\ee
We have also used the resummation method of Ref. \cite{francesi1980}, 
which takes into account the singularity structure of the Borel transform
of the perturbative series. We obtain \cite{footnote3}
\be
\widetilde{f}_\infty \approx - 0.089 \pm 0.017, \qquad \qquad
f_0 \approx 0.015 \pm 0.002.
\label{eq3.25}
\ee
The ratio $\widetilde{f}_\infty/f_0$ can be determined precisely using
Eq. \reff{relf0ftildeinf} and the precise Monte Carlo determinations of 
the critical exponents:
\be
{\widetilde{f}_\infty\over f_0} = -4.883 \pm 0.017.
\ee
We can use this result to test the accuracy of the estimates \reff{eq3.24}
and \reff{eq3.25}.
Using Eq. \reff{eq3.24} we obtain $\widetilde{f}_\infty/f_0=-3.16$, 
while Eq. \reff{eq3.25} gives $\widetilde{f}_\infty/f_0=-5.9 \pm 1.4$. 
The estimates have an error of approximately 20-30\%. 
It is reassuring that the errors obtained with the resummation 
method correctly describe the discrepancy.

\section{Asymptotic behaviour via Laplace-de Gennes transform} \label{sec4}

Improved estimates of the asymptotic behaviour of $f(\rho)$
can be obtained from the
precise results that have been derived for $O(N)$ spin models, using 
the Laplace-de Gennes transform method 
\cite{DeGennes_72,DesCloizeaux_74_80,DesCloizeaux-Jannink_book}.
We introduce in 
$d$ dimensions the two-point function 
\be
G({\bf r};\beta) =\, \sum_{N=0}^\infty e^{-\beta N} c_N({\bf r}),
\ee
its Fourier transform 
\be
\widetilde{G}({\bf q};\beta) = \sum_r e^{i{\bf q}\cdot {\bf r}}
        G({\bf r};\beta),
\ee
and the correlation length
\be
\xi^2 (\beta) \equiv {1\over 2d}{\sum_r |{\bf r}|^2 G({\bf r};\beta)\over
                 \sum_r  G({\bf r};\beta) }\;.
\ee
In the critical limit $G({\bf r};\beta)$ and $\widetilde{G}({\bf q};\beta)$
have the scaling form
\cite{Fisher-Burford_67,Fisher-Aharony_73_74}
\be
{G({\bf r};\beta)\over \sum_{\bf r} G({\bf r};\beta)} \approx 
 {1\over \xi^d(\beta)} D(\rho), \qquad\qquad
{\widetilde{G}({\bf q};\beta)\over \widetilde{G}({\bf 0};\beta)}
\approx  \widetilde{D}(Q),
\ee
where $\rho = r/\xi (\beta)$ and  $Q \equiv |{\bf q}| \xi (\beta) $.
The function $\widetilde{D}(Q)$ is the $d$-dimensional 
Fourier transform of $D(\rho)$.

Several facts are known about the function $\widetilde{D}(Q)$. For $Q\ll 1$,
$\widetilde{D}(Q)$ has a regular expansion in terms of $Q^2$ as 
\be
\widetilde{D}^{-1}(Q) = 1 + Q^2 + \sum_{i=2}^\infty b_n Q^{2 n}. 
\label{DsmallQ}
\ee
The coefficients $b_n$, $n=2,3,4,5$ have been computed \cite{Bray_76}
in the $\epsilon$-expansion up to $O(\epsilon^4)$ and 
\cite{Campostrini-etal_98} in the 
fixed-dimension expansion in $d=3$ up to $O(g^5)$. 
It turns out they are extremely small
and satisfy $b_2 \gg b_3 \gg b_4 \ldots $. For $b_2$ the 
explicit formulae are 
\bea
b_2 &=& - 0.000235007 \epsilon^2 [1 + 1.0632 \epsilon + O(\epsilon^2)], \\
b_2 &=& - 0.00015432 g^2 [1 + 0.0780213 g +  0.0465896 g^2 + O(g^3)].
\eea
Here $g$ is the renormalized four-point coupling \cite{footnote4} 
constant whose
critical value 
\cite{Baker,francesi1980,Nickel-Muthukumar,Nickel_1991,Pelissetto-Vicari_98}, 
in the normalization we use, is $g^{*} \approx 1.39$. 
The expansions for $b_2$, $b_3$, $\ldots$, can be resummed using the 
method of Ref. \cite{francesi1980}, obtaining
\cite{Campostrini-etal_98} 
$b_2 \approx (-3\pm 1)\cdot 10^{-4}$.
Analogously $b_3 \sim 10^{-5}$.
An exact-enumeration study confirmed these results
and gave the bound $-3\cdot 10^{-4} \ltapprox b_2 \ltapprox 0$. 
In two dimensions estimates have been obtained from the analysis 
\cite{Campostrini-etal_96} of exact-enumeration expansions
on the triangular, square and honeycomb
lattices: $b_2 \approx 0.00015 (20)$, $|b_3|\ltapprox 3 \cdot 10^{-5}$.
In the following we will not need the explicit values of the 
constants $b_i$. Indeed they are too small to give any numerically important 
effect. Thus, for $Q\to 0$, we can approximate $\widetilde{D}^{-1}(Q)$ with 
$1 + Q^2$.

For $Q\gg 1$, the behaviour of $\widetilde{D}(Q)$ is predicted 
by a short-distance  renormalization-group analysis 
\cite{Fisher-Langer_68,BAJ} and one has
\be
\widetilde{D}(Q) = 
       {D_0\over Q^{2-\eta}} + {D_1\over Q^{2-\eta+(1-\alpha)/\nu}} + 
       {D_2\over Q^{2-\eta+1/\nu}} + \ldots
\label{DlargeQ}
\ee
The exponents $\eta$ and $\alpha$ are related to the 
exponents $\gamma$ and $\nu$ by
\bea
\eta &=& 2 - {\gamma\over \nu}, \\
\alpha &=& 2 - d \nu.
\eea
The constants $D_0$, $D_1$ and $D_2$ have been computed in the 
$\epsilon$-expansion \cite{Fisher-Aharony_73_74,Bray_76}:
\bea
D_0 &=& 1 - 0.0317391 \epsilon^2 - 0.0353978 \epsilon^3 + 
    O(\epsilon^4) \\
D_1 &=& {1\over2}\left( 1 + {5\over 8}\epsilon - 0.4822 \epsilon^2\right)  + 
          D_{13} \epsilon^3 + O(\epsilon^4),\\
D_2 &=&- {3\over2} \left( 1 + {5\over 24}\epsilon - 0.17653 \epsilon^2\right) 
           + D_{23} \epsilon^3 + O(\epsilon^4).
\eea
The coefficients
$D_{13}$ and $D_{23}$ are not known but satisfy the  relation:
\be
  D_{13} + D_{23} =\, 0.05934.
\ee
Resumming the perturbative series, we obtain
\be
D_0 \approx 0.97\pm 0.02, \qquad
D_1 \approx 0.71\pm 0.04, \qquad
D_2 \approx - 1.70\pm 0.06. 
\ee
Using the longer series for $D_1+D_2$ we obtain $D_1+D_2\approx -0.97\pm 0.02$.
It has been remarked by Aharony and Fisher 
\cite{Fisher-Aharony_73_74} that one can also
rewrite 
\be
D_1 = {1\over \alpha} \left(\gamma - 1 + O(\epsilon^3)\right)\; ,
\ee
and thus, for $d=3$, using the known values of the critical exponents, we
obtain a similar estimate $D_1 \approx 0.67$.

Finally one can determine the large-$\rho$ behaviour of $D(\rho)$.
For $\rho\to\infty$, using the notations of Ref. \cite{Campostrini-etal_98},
we have
\be
D(\rho) \approx \, 
A^* \rho^{-(d-1)/2}
    e^{- \sqrt{S_M^*} \rho} \; ,
\label{Dlarger}
\ee
where $A^*$ and $S_M^*$ are non-perturbative constants.
The constant $A^*$ can be related to the residue $S^*_Z$ of the propagator 
at the mass pole. It is  given by
\be
S^{*}_Z =\, {1\over2} \left. {\partial^2\over \partial Q^2} 
    {D}^{-1} (Q) \right |_{Q^2 = - S^{*}_M}\; .
\ee
One obtains
\be
A^{*} = {1\over 2 S^{*}_Z} (2 \pi)^{(1-d)/2} \; 
       \left(S^{*}_M\right)^{(d-3)/4}.
\ee
The constants $S_Z^{*}$ and $S_M^{*}$ have been computed 
\cite{Campostrini-etal_98}
in the $\epsilon$-expansion up to $O(\epsilon^4)$ and in the expansion in 
fixed dimension $d=3$ up to $O(g^4)$:
\bea 
S_Z^{*} &=& 1 + 0.000489100 \epsilon^2 + 0.000522425 \epsilon^3 + 
          O(\epsilon^4),\\
S_M^{*} &=& 1 - 0.000241274 \epsilon^2 - 0.000257303 \epsilon^3 +
          O(\epsilon^4),\\
S_Z^{*} &=& 1 + 0.000335466 g^2 + 0.0000140322 g^3 + O(g^4), \\
S_M^{*} &=& 1 - 0.000163057 g^2 - 0.0000088348 g^3 + O(g^4).
\eea
It is evident from these expansions that both constants are 
one with very small corrections. Resumming the expansions, we obtain in three
dimensions
$S_M^{*} \approx 1 - (3\pm 1) \cdot 10^{-4}$ and 
$S_Z^{*} \approx 1 + (5\pm 1) \cdot 10^{-4}$. In two dimensions 
estimates have been obtained from the analysis of 
exact-enumeration expansions \cite{Campostrini-etal_96}:
$S_M^{*} \approx  1 + (1\pm 2) \cdot 10^{-4}$,
$S_Z^{*} \approx  1 - (2\pm 4) \cdot 10^{-4}$.

{}From the asymptotic behaviour of ${D}(\rho)$ 
we obtain corresponding predictions for $f(\rho)$, using
the fact that \cite{DesCloizeaux_74_80,DesCloizeaux-Jannink_book}
\be
c_N({\bf r}) =\, {1\over 2\pi i} \int^{+i\infty}_{-i\infty}
   d\beta \, e^{\beta N} G({\bf r};\beta)\; .
\ee
We report here the results; the derivations can be found 
e.g. in \cite{DesCloizeaux_74_80,DesCloizeaux-Jannink_book}.

We begin by computing the large-$Q$ behaviour of $\widetilde{f}(Q)$.
Using Eq. \reff{DlargeQ}
we obtain the expression \reff{ftildelargeQ} with
\be
\widetilde{f}_\infty = {1\over \pi}
   \Gamma(\gamma) D_1 
     \sin(\pi d \nu) \Gamma(d \nu) 
    \left( {\Gamma(\gamma)\over \Gamma(\gamma + 2 \nu)}
     \right)^{(\theta + d)/2}  \; .
\ee
Using the numerical values of the exponents, we obtain in two dimensions 
\bea
\theta &=& {11\over 24},\\
{\widetilde{f}_\infty\over D_1} &=& -0.11062768\ldots
\eea
and in three dimensions
\bea
\theta &=& 0.2680 \pm 0.0011,\\
{\widetilde{f}_\infty\over D_1} &=& -0.12393\pm0.00026.
\eea
Using Eq. \reff{relf0ftildeinf}, we obtain correspondingly
\be
{f_0\over D_1} = \cases{ 0.050548...... & for $d=2$, \cr 
                         0.02539\pm 0.00014 & for $d=3$.} 
\ee
In two dimensions we do not have any prediction for $D_1$, but, on the 
basis of the $\epsilon$-expansion result, we expect $D_1$ to be of 
order 1, so that a reasonable guess is 
$-0.05\ltapprox \widetilde{f}_\infty\ltapprox -0.15$
and $0.02 \ltapprox f_0 \ltapprox  0.07$.
In three dimensions, using $D_1\approx 0.71\pm 0.04$,
we obtain $\widetilde{f}_\infty \approx -0.088\pm 0.006$  and 
$f_0 \approx 0.018\pm 0.001$. These estimates are in good agreement with the 
results obtained in Sec. \ref{sec3}.

It is interesting to compare
the small-$\rho$ behaviour of $f_{\rm ph}(\rho)$ with that of 
the exact function $f(\rho)$.
Using Eq. \reff{frhoapprox}, we obtain 
$f_{0,\rm ph} \approx f_{\rm ph} 2^\theta \approx 0.0193$, 
which does not differ significantly from the estimate of $f_0$ reported above.
We can also compare the two predictions within the 
$\epsilon$-expansion. We have
\be
{f_0\over f_{\rm ph} 2^\theta} = 
  1 + 0.109663 \epsilon^2 + O(\epsilon^3),
\ee
which shows that the phenomenological approximation is 
essentially correct, with an expected discrepancy of order 10\%.

Using Eq. \reff{Dlarger}, one obtains the asymptotic 
behaviour \reff{flargerho}
where $f_\infty$ and $D$ are given by 
\bea
   D &=&  {1-\nu\over \nu}\left( \nu^2 S^{*}_M 
         {\Gamma(\gamma)\over \Gamma(\gamma + 2 \nu)}\right)^{\delta/2} 
   \label{Dconstant} \\
   f_\infty &=& {\left( S^{*}_M \right)^{(d-3)/4} \over 2 S^{*}_Z} 
    {\Gamma(\gamma)\over (2\pi)^{d/2} (1-\nu)^{1/2}} 
   \left( \nu^2 S^{*}_M 
         \right)^{\delta(\nu(d+1) - 2 \gamma + 1)/4}
   \left({\Gamma(\gamma)\over \Gamma(\gamma + 2 \nu)}
          \right)^{\delta(2 - 2 \gamma + d)/4}
    \label{finftyconstant}
\eea
Using the values of 
$S^{*}_M$ and $S^{*}_Z$ we have reported before and the values of the 
exponents $\gamma$ and $\nu$ we get in two dimensions:
\bea
\delta &=& 4, \\
\sigma &=& {5\over 8}, \\
D &=& 0.02771 \pm 0.00001, \\
f_\infty &=& 0.04273 \pm 0.00002.
\eea
In three dimensions, using
$\nu = 0.58758 \pm 0.00007$ \cite{Belohorec-Nickel_97} and
our result \cite{Caracciolo-etal_97} $\gamma = 1.1575 \pm 0.0006$,
we obtain
\bea
\delta &=& 2.4247 \pm 0.0004,\label{delta-d3}\\
\sigma &=& 0.255 \pm 0.002,\label{sigma-d3}\\
D &=&  0.1434 \pm 0.0002,\label{D-d3}\\
f_\infty &=&  0.01581 \pm 0.00002.\label{f-d3}
\eea
Notice that the estimates of $D$ and $f_\infty$ would not have 
significantly changed, had we used the Gaussian values 
$S^{*}_M = S^{*}_Z = 1$. In three dimensions the error is 
dominated by the error on $\gamma$ and $\nu$.

We can also use Eqs. \reff{Dconstant} and \reff{finftyconstant}
to derive $\epsilon$- and $g$-expansions for $D$ and $f_\infty$. 
We obtain in $4-\epsilon$ dimensions
\bea
D &=& {1\over4} - 0.0877837 \epsilon - 0.0327168 \epsilon^2 + 
      0.0394476 \epsilon^3 + O(\epsilon^4), \\
f_\infty &=& {1\over 16\pi^2} + 0.00579036 \epsilon + 0.00223513 \epsilon^2 + 
      0.00152100 \epsilon^3 + O(\epsilon^4);
\label{finfinityepsilon}
\eea
in fixed dimension $d=3$ we get
\bea
\hskip -1truecm
D &=&  {1\over 4} - 0.0877837 g + 0.0153578 g^2 - 0.0104654 g^3 + O(g^4), \\
\hskip -1truecm
f_\infty &=&  {1\over (2\pi)^{3/2}} - 0.00534423 g + 0.000981669 g^2 -
        0.000734351 g^3 + O(g^4).
\eea
Resumming the expansions using the method of Ref. \cite{francesi1980},
we obtain in the two cases: 
the $\epsilon$-expansion gives $D = 0.1461 \pm 0.0040$, 
$f_\infty = 0.0138 \pm 0.006$, while the $g$-expansion
gives $D = 0.1445 \pm 0.0015$, $f_\infty = 0.01602 \pm 0.00010$.
These results are less accurate than the previous ones, that were
obtained using the precise Monte Carlo estimates of the critical
exponents. Note that the estimate of $f_\infty$
obtained using the $\epsilon$-expansion is not compatible, with the 
quoted error bars, with the estimate \reff{f-d3}. This is not surprising
since in the expansion (\ref{finfinityepsilon}) all coefficients
have the same sign: therefore, a Borel resummation based on the
large-order behaviour of the coefficients (that predicts
coefficients alternating in sign) is not expected to work well.

It is interesting to compare the estimates 
\reff{D-d3} and \reff{f-d3} with the phenomenological 
approximation \reff{Dph-d3} and \reff{fph-d3}. It is remarkable that 
the discrepancy is so tiny, precisely of 
0.8\% for $D$ and of 1.1\% for $f_\infty$. 
Also in two dimensions the phenomenological approximation 
works reasonably well: 
the discrepancy is of 5\% for $D$ and of 9\% for $f_\infty$.
This nice
agreement can be understood within the $\epsilon$-expansion.
Indeed
\bea
{D\over D_{\rm ph}} &=& 1 - 0.012109 \epsilon^2 + 
                            0.0039898 \epsilon^3 + O(\epsilon^4),
\\
{f_\infty\over f_{\rm ph}} &=& 1 - 0.0083917 \epsilon^2 
                   - 0.0065622 \epsilon^3 + O(\epsilon^4).
\eea
The $\epsilon$-expansions of $D$ and $f_\infty$ and of their 
phenomenological approximations $D_{\rm ph}$ and $f_{\rm ph}$ 
differ by terms that are very small. Setting $\epsilon = 1$,
one finds an expected difference of order 1\%, in agreement
with the estimates above.

Notice that also the exponent $\sigma$ does not differ
significantly from $\theta$ in three dimensions. 
This explains the success of the phenomenological approximation
\reff{frhoapprox} 
for $\rho\to\infty$. Indeed, in this limit, we have
\be
{f_{\rm ph}(\rho)\over f(\rho)} \approx\, 
  {f_{\rm ph}\over f_\infty} \rho^{\theta - \sigma}
  \exp\left[ - (D_{\rm ph} - D)\rho^\delta\right] \approx \,
  1.011\, \rho^{0.013} \exp(- 0.0013 \rho^\delta),
\ee
so that 
\be
{f(\rho) - f_{\rm ph}(\rho)\over f(\rho)} \approx 
\cases{ {\rm 1\%} & for $\rho = 2$, \cr
        {\rm 3\%} & for $\rho = 5$, \cr
        {\rm 26\%}& for $\rho = 10$.}
\ee
However $f(5) \approx 2\cdot 10^{-5}$, and $f(10) \approx 8 \cdot 10^{-19}$, 
so that in practice $f(\rho)$ can be sampled up to $\rho\approx 5$--6.
Therefore, in the region accessible to numerical simulations,
$f_{\rm ph}(\rho)$ provides an accurate description of the large-$\rho$
behaviour of the EEDF.

Finally, we can use \reff{DsmallQ} to get predictions for the moments
of $f(\rho)$. It is simple to show that the invariant ratios
\be
M_{2k} = {\sum_r |{\bf r}|^{2k} P_N({\bf r}) \over 
         \left(\sum_r |{\bf r}|^{2} P_N({\bf r})\right)^k}
\ee
approach, for $N\to\infty$, universal constants $M^{*}_{2k}$ given by
\be
M^{*}_{2k} = 
{\Gamma(\gamma + 2 \nu)^k \over \Gamma(\gamma + 2 k \nu) \Gamma(\gamma)^{k-1}}
\, \left[ 1 - b_2 (k-1) \right] k! 
\prod_{j=0}^{k-1} \left(1 + {2j\over d}\right),
\label{Mstar}
\ee
where, we have neglected all $b_n$ with 
$n\ge 3$ and all powers of $b_2$. Notice that Eq. \reff{Mstar} is exact 
for $k=2$.

Again, we can compare the exact expression \reff{Mstar} with the prediction 
obtained by using the phenomenological representation \reff{frhoapprox}:
\be
M_{{\rm ph},2k}^*=\, \Gamma\left({2k+\theta+d\over\delta}\right)
                     \Gamma\left({\theta+d\over\delta}\right)^{k-1}
                     \Gamma\left({2+\theta+d\over\delta}\right)^{-k}.
\label{Mstarapprox}
\ee
Numerically we have in three dimensions:
\begin{eqnarray}
M^*_4 = 1.51397(79) && \qquad M_{{\rm ph},4}^* = 1.50876(23), \\
M^*_6 = 3.018(4)    && \qquad M_{{\rm ph},6}^* = 2.993(1),    \\
M^*_8 = 7.392(15)   && \qquad M_{{\rm ph},8}^* = 7.292(5),    \\
M^*_{10} = 21.35(6) && \qquad M_{{\rm ph},10}^* = 20.94(2).    
\end{eqnarray}
The phenomenological predictions are definitely not exact, but they
show very small differences with respect
to the exact ones. Since the moments define uniquely the distribution function,
this implies that Eq. \reff{frhoapprox} is not only a good approximation 
in the regions of large and small values of $\rho$, but that it also provides 
a good overall parametrization of the EEDF.

We can also consider the ``wall-to-wall" distribution $P_{N,w}(x)$.
It is easy to see that for large $\rho$ we have Eq. \reff{Eq2.27} with 
\bea
 \sigma_w &=& {\delta}\left(\nu - \gamma + {1\over2}\right), \\
 f_{w,\infty} &=& f_\infty \left({2 \pi \nu } \right)^{(d-1)/2}
    \left( {\nu^2 S^{*}_M \Gamma(\gamma)\over \Gamma(\gamma + 2 \nu)}
         \right)^{\delta (1-d)/4}.
\eea
In two dimensions we obtain
\bea
 \sigma_w &=& - {3\over8}, \\
 f_{w,\infty} &=& 0.32167 \pm 0.00017,
\eea
and in three dimensions 
\bea
 \sigma_w &=& - 0.1695 \pm 0.0016, \label{sigmaw-d3} \\
 f_{w,\infty} &=& 0.2855 \pm 0.0004. \label{fw-d3}
\eea

\section{Monte Carlo study of the distribution function}

The EEDF has been extensively studied numerically in three
dimensions. 
The Monte Carlo work essentially 
focused on the exponents and verified that the data could be well
described by the phenomenological expression \reff{frhoapprox}.
Baumg\"artner \cite{Baumgartner_81} computed the exponent $\delta$ obtaining
\be
\delta = 2.44 \pm 0.05,
\ee
in very good agreement with the theoretical result 
\reff{delta-d3}. 
The exponent $\theta$ appearing in the phenomenological expression
\reff{frhoapprox} was computed by several 
groups, obtaining
\be
\theta=\cases{0.270\pm 0.006 & Ref. \protect\cite{Baumgartner_81}, \cr
              0.27           & Ref. \protect\cite{Dayantis-Palierne_91}, \cr
              0.262\pm 0.013 & Ref. \protect\cite{Eizenberg-Klafter_93}, \cr
              0.224\pm 0.006 & Ref. \protect\cite{Pedersen-etal_96}.
              }
\ee
All estimates but the last one do not differ sensibly from our theoretical 
results for $\sigma$ and $\theta$, and as expected, they lie between 
these two estimates. The estimate of Ref. \cite{Pedersen-etal_96} is instead
too low; probably, the numerical data are affected by large corrections to 
scaling.

In this Section we want to extend these numerical analyses, 
checking the renor\-ma\-li\-zation-group predictions presented in the 
previous Sections. We will not use the phenomenological expression
\reff{frhoapprox}, but we will compute the exponents 
$\delta$, $\sigma$, and $\theta$ studying the large-$\rho$ and 
small-$\rho$ behaviour of the EEDF. At the same time we will 
be able to compute the constants $f_0$, $f_\infty$ and $D$ and to 
compare them with the theoretical predictions. 

In order to compute the EEDF, we have generated $N$-step self-avoiding walks 
on a three dimensional cubic lattice, using the pivot algorithm 
\cite{Lal_69,MacDonald-etal_85,Madras-Sokal_88,Li-etal_95}. 
Since in three dimensions corrections to scaling are particularly strong,       
we generated long walks with $500\le N\le 32000$.

First of all, we have checked the prediction \reff{Mstar} for 
the invariant ratios
$M_{2k}$, for $k=2,3,4,5$. In Table \ref{Mratios} we report the Monte Carlo 
estimates of $M_{2k}$ for various values of $N$. We have 
performed the extrapolation to $N\to\infty$ using
\be
M_{2k} = M_{2k}^* + B_{2k} \left({8000\over N}\right)^\Delta\; .
\ee
The final estimates are compatible with the less precise results 
of Ref. \cite{Bishop-Clarke_91} and are in very good agreement with the 
theoretical predictions. The corrections to scaling appear 
to be quite important: we estimate 
$\Delta \approx 0.55 \pm 0.08$, in agreement with the 
renormalization-group and Monte Carlo predictions \cite{footnote5}.
Note that, at our level of precision, the phenomenological predictions
\reff{Mstarapprox} are not consistent with the numerical data.

It is interesting to observe that we can use Eq. \reff{Mstar} 
to obtain independent estimates of the critical exponents. 
For instance, using $b_2 = - (3\pm 1) \cdot 10^{-4}$ 
\cite{Campostrini-etal_98},
$\nu = 0.58758\pm 0.00007$ \cite{Belohorec-Nickel_97}, 
and the Monte Carlo result for $M^*_4$, we obtain
\be
\gamma = 1.1576 \pm0.0013,
\ee
which is in perfect agreement 
with the more precise estimate of Ref. \cite{Caracciolo-etal_97}.
It should be noticed that most of the error on $\gamma$ is due to the 
error on $M^*_4$, and, indeed, with the existing estimates of $\nu$ and 
$b_2$, it would be possible to obtain with this method an estimate 
of $\gamma$ as precise as that given in Ref. \cite{Caracciolo-etal_97}.
By considering $M^*_4$ and $M^*_6$ one can try to estimate 
$\nu$ and $\gamma$ simultaneously. One obtains
$\nu = 0.588(6)$, $\gamma = 1.159(24)$. 
Finally, using the estimates of $\gamma$ and $\nu$ and our Monte Carlo
result for $M^*_4$, we obtain a bound on $b_2$. We get
$|b_2|<1.4\cdot 10^{-3}$.

In addition to $M_{2k}$, one can consider the invariant ratios 
\be
K_{2k} = {\sum_r (x^{2k} + y^{2k} + z^{2k}) P_N({\bf r}) \over 
          \left[ \sum_r (x^{2} + y^{2} + z^{2}) P_N({\bf r}) \right]^k }.
\ee
The rotational invariance of the critical limit gives 
\be
R_{2k} = {K_{2k}\over M_{2k}} \to {3\over 2k + 1}.
\ee
In Table \ref{Rratios} we report the 
estimates of the ratios $R_{2k}$ for $k=2,3,4,5$ and several
values of $N$. Notice that, in this case, 
corrections to scaling are practically
absent. This is in agreement with the analysis of Ref.
\cite{Campostrini-etal_98}
which showed that quantities like $R_{2k}$ have corrections of the 
form $N^{-\tau}$ with $\tau \approx 2\nu$. Thus, they are much smaller than the 
standard corrections which behave as $N^{- \Delta}\sim N^{-0.5}$. 

Let us now consider the EEDF itself. The 
computation of $P_{N}(x)$ from the Monte Carlo data is straightforward, but
it is less clear how to estimate the 
error bars. If the walks are generated independently, and
$\hat{p}(x)$ is the estimate of $P_{N}(x)$, the error is 
\be
\left[ {1\over N_{MC}} \hat{p}(x) (1 - \hat{p}(x)) \right]^{1\over2},
\label{errors}
\ee
where $N_{MC}$ is the number of Monte Carlo iterations. In our 
case, however, the walks are not generated independently. Therefore,
one should take into account the autocorrelation time of the algorithm
and the fact that estimates at different
values of $x$ are correlated. In practice it is not feasible 
to take into account all these effects. 
We have simply observed that since global observables decorrelate
after a few accepted
pivot moves, a reasonable estimate of the errors can be obtained
by replacing  in Eq. \reff{errors} $N_{MC}$ with $f_N N_{MC}$, 
where $f_N$ is the 
acceptance fraction of the algorithm. Correlations between different points are
neglected.

\begin{figure}
\begin{center}
\epsfxsize = 0.8 \textwidth
\epsfbox{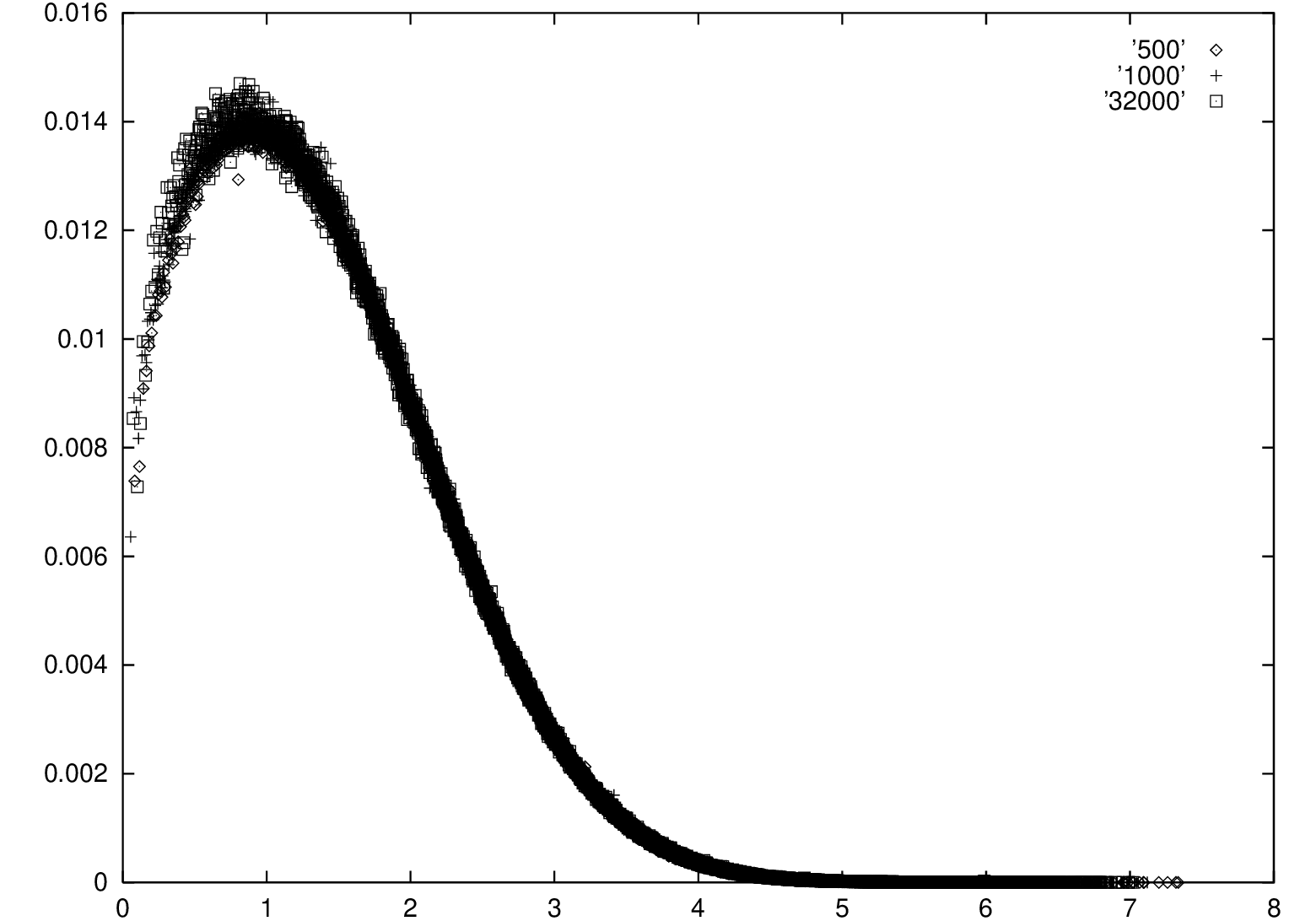}  
\end{center}
\caption{ Plot of $f(\rho)$ for different values of $N$.
 } \label{figura1}
\end{figure}
\begin{figure}
\begin{center}
\epsfxsize = 0.8 \textwidth
\epsfbox{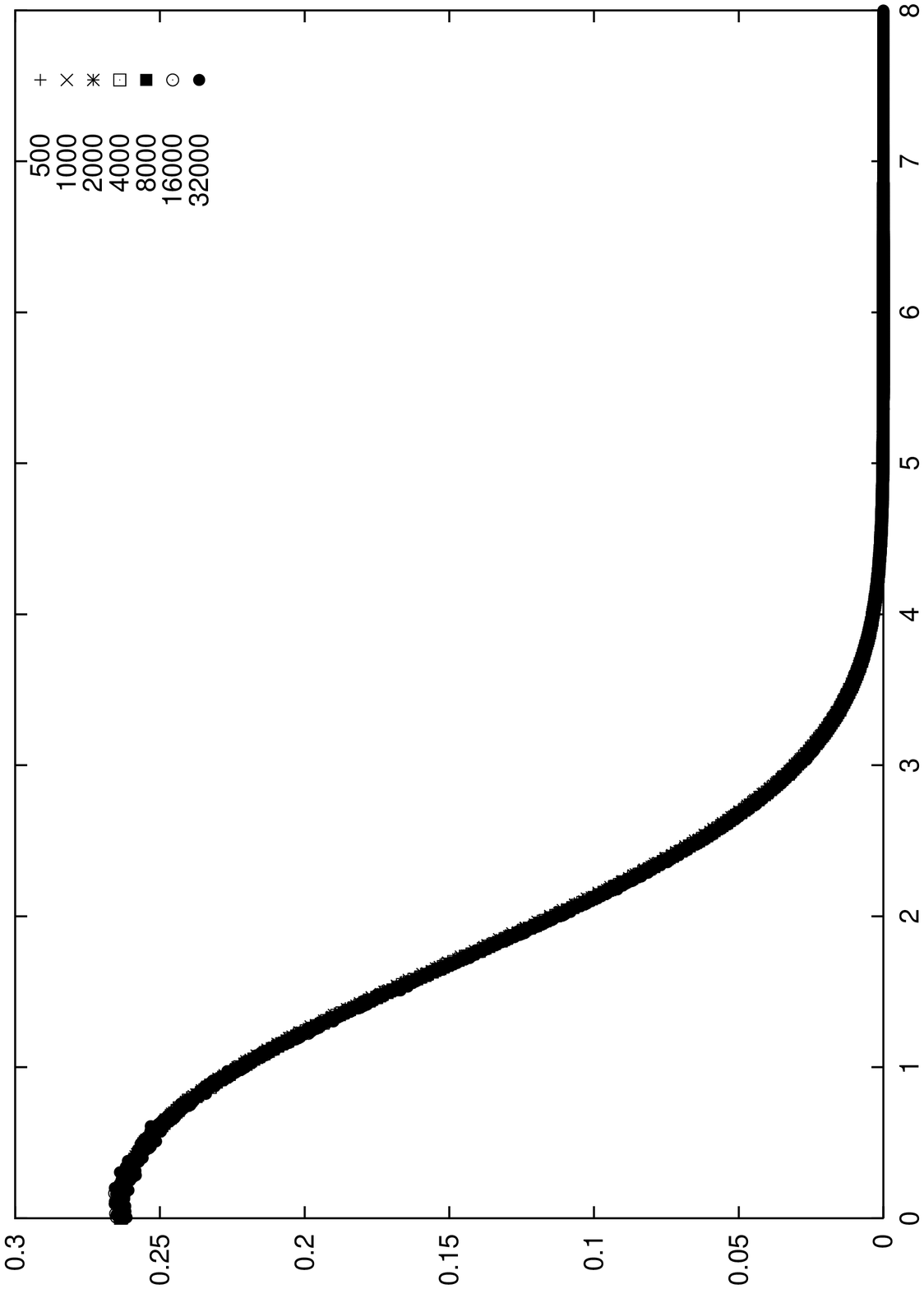}  
\end{center}
\caption{ Plot of $f_w(\rho)$  for different values of $N$.
 } \label{figura2}
\end{figure}

The functions $f(\rho)$ and $f_w(\rho)$ are reported in Figs. 
\ref{figura1} and \ref{figura2}. The data fall on a single
curve as expected: within the accuracy of the plot, no corrections to scaling
are visible, but, as we shall discuss later, corrections are present
if one looks at the data in more detail.

Let us now study the asymptotic behaviour of the EEDF. We will begin by
considering the wall-to-wall EEDF $f_w(\rho)$. In order to study
its large-$\rho$ behaviour, we have performed two different sets of fits:
\bea
\log (f_w(\rho)) &=& \log f_{w,\infty} - D \rho^\delta, 
\label{fit11-wall}\\
\log (\rho^{0.169}\ f_w(\rho)) &=& \log f_{w,\infty} - D \rho^\delta.
\label{fit12-wall}
\eea
In the first fit we have neglected 
the power term $\rho^{\sigma_w}$ that appears in the asymptotic 
behaviour of $f_w(\rho)$, in the second one we have taken this term
into account using the theoretical prediction for $\sigma_w$,
Eq. (\ref{sigmaw-d3}). For $\rho\to\infty$, both fits should give the 
correct result for $D$ and $\delta$, while only the second one 
gives an estimate of $f_{w,\infty}$. 
There are two types of systematic errors in these fits.
First, there are scaling corrections: the scaling curve is obtained
only in the limit $N\to\infty$. Secondly, there are non-asymptotic corrections:
Eqs.  \reff{fit11-wall} and \reff{fit12-wall} are valid only 
asymptotically for $\rho\to\infty$. In order to detect scaling corrections,
we have performed fits at fixed values of $N$. Then, we have compared the 
results, looking for systematic variations of the estimates with the 
length of the walk. The final result is obtained including in the fit
only walks with $N\ge N_{\rm min}$, where $N_{\rm min}$ is chosen so that 
the estimates for all $N\ge N_{\rm min}$ are independent of $N$ 
within error bars. A similar strategy has been used to detect non-asymptotic 
effects: we have performed several fits using in each case only
data with $\rho\ge \rho_{\rm min}$. Looking at the variation of the 
estimates with $\rho_{\rm min}$, we can estimate the non-asymptotic 
corrections.
The results of the fits for fixed values of $N$ 
and for three different values of $\rho_{\rm min}$
are reported in Table \ref{wall-fit1}. Apparently
they do not show any systematic dependence on 
$N$, except perhaps $N=500$ and $N=1000$: 
indeed the estimate of $\delta$ for $N=500$, $1000$
are slightly higher than the estimates obtained for larger values 
of $N$, while the estimates of $D$ are slightly smaller. One may
suspect that these results are affected by scaling corrections of size
comparable with the statistical error. For this reason, our final 
estimates are obtained using all data with $N\ge 2000$ only.
We obtain from the first fit
\bea
\delta &=& \cases{2.413\pm 0.006 & \hskip 1truecm $\rho_{\rm min} = 3.0$, \cr
                  2.420\pm 0.014 & \hskip 1truecm $\rho_{\rm min} = 3.5$,} 
\\ 
D &=& \cases{0.150 \pm 0.002 & \hskip 1truecm $\rho_{\rm min} = 3.0$, \cr
             0.148 \pm 0.004 & \hskip 1truecm $\rho_{\rm min} = 3.5$.}
\eea
The second fit gives
\bea
\delta &=& \cases{2.458\pm 0.006 & \hskip 1truecm $\rho_{\rm min} = 3.0$, \cr
                  2.455\pm 0.013 & \hskip 1truecm $\rho_{\rm min} = 3.5$,} 
\\ 
D &=& \cases{0.136 \pm 0.002 & \hskip 1truecm $\rho_{\rm min} = 3.0$, \cr
             0.137 \pm 0.004 & \hskip 1truecm $\rho_{\rm min} = 3.5$.}
\eea
These results are in good agreement with the theoretical estimates 
reported above, cf. Eqs. (\ref{delta-d3}) and (\ref{D-d3}).  
Indeed, increasing $\rho_{\rm min}$, we observe the expected convergence
to the theoretical results.
The second fit gives also an estimate of $f_{w,\infty}$:
we find
\beq
f_{w,\infty} = \cases{
       0.274 \pm 0.002 & \hskip 1truecm $\rho_{\rm min} = 3.0$, \cr
       0.276 \pm 0.006 & \hskip 1truecm $\rho_{\rm min} = 3.5$,}
\eeq
which converges to the estimate (\ref{fw-d3}) for large 
values of $\rho_{\rm min}$.

In order to estimate $\sigma_w$ and $f_{w,\infty}$,
we have performed fits of the form
\beq
\log\left[ f_w(\rho) \exp(D \rho^\delta)\right] = 
\log f_{w,\infty} + \sigma_w \log \rho,
\eeq
using the theoretical estimates of $D$ and $\delta$, for various 
values of $\rho\ge \rho_{\rm min}$. These fits are extremely 
unstable. Indeed the EEDF drops rapidly to zero 
(see Fig. \ref{figura2}) so that the fit uses data in a small 
interval in which $\log \rho$ does not vary significantly. 
Results with reasonable errors can be obtained only for 
$\rho_{\rm min} \ltapprox 3$, and thus we have analyzed the data with  
$\rho_{\rm min} = 2,2.5$ and $3$. The results for fixed values of $N$
are reported in Table \ref{wall-fit2}. 
Looking at the table, one immediately sees that there are strong 
non-asymptotic corrections. Clearly the asymptotic behaviour 
sets in only for very large values of $\rho$. Looking at the 
data with $\rho_{\rm min} = 3.0$, one sees that the estimates show 
a systematic trend as $N$ increases. If we analyze together all
data with $N\ge N_{\rm min}$ we have ($\rho_{\rm min} = 3.0$) 
\bea
\sigma_w &=& \cases{
    - 0.184 \pm 0.004 & \hskip 1truecm $N_{\rm min} =  1000$, \cr
    - 0.173 \pm 0.005 & \hskip 1truecm $N_{\rm min} =  2000$, \cr
    - 0.169 \pm 0.005 & \hskip 1truecm $N_{\rm min} =  4000$, \cr
    - 0.157 \pm 0.006 & \hskip 1truecm $N_{\rm min} =  8000$, }
\\
f_{w,\infty} &=& \cases{
      0.289 \pm 0.002 & \hskip 1truecm $N_{\rm min} =  1000$, \cr
      0.284 \pm 0.002 & \hskip 1truecm $N_{\rm min} =  2000$, \cr
      0.282 \pm 0.002 & \hskip 1truecm $N_{\rm min} =  4000$, \cr
      0.278 \pm 0.002 & \hskip 1truecm $N_{\rm min} =  8000$.}
\eea
The presence of confluent corrections and of non-asymptotic 
terms of opposite sign makes difficult to evaluate reliably 
$\sigma_w$ and $f_{w,\infty}$: clearly large values of $N$ are needed
to see the scaling regime and large values of $\rho$
are required to observe the correct asymptotic behaviour. From the 
results of the fits above we can conclude that there is a reasonable agreement 
between the theoretical estimates and the numerical results 
although a precise quantitative check is difficult.

Let us now consider the distribution function $f(\rho)$. In order to 
determine this function, we have computed 
the probability $P_n({\bf r})$ from the Monte Carlo data.
A graph of this quantity as a function of $r^2$
shows strong oscillations due to the underlying lattice structure.
In order to reduce these effects, we have used a procedure analogous 
to that used in Ref. \cite{Dayantis-Palierne_91}.
Given a number $N_{sh}$, we define $r^2_n = n N_{sh}$ and an averaged
distribution function
\beq
P_{N}^{av} (r_n) = \, 
   {1\over N_n} \sum_{{\bf r}: r^2_{n-1} < r^2 \le r^2_n} P_N({\bf r}),
\label{Pav}
\eeq
where $N_n$ is the number of lattice points in the 
shell $r^2_{n-1} < r^2 \le r^2_n$. 
For $N_{sh}$ fixed, in the scaling limit 
$|{\bf r}|\to \infty$, $N\to\infty$, with $\rho$ fixed, 
$P_{N}^{av} (r)$ converges to $f(\rho)$, so that one can use 
the distribution \reff{Pav} in order to compute the EEDF. 
The advantage is that lattice oscillations disappear in the 
averaging procedure. Of course, one should always check 
that the results do not depend on $N_{sh}$. As expected, as long
as the number of points falling in each shell is sufficiently large
and $\sqrt{N_{sh}} \ll \xi_N$, the final estimates are not sensitive
to $N_{sh}$.

We have closely repeated the analysis performed for the 
wall-to-wall EEDF. The final results are in reasonable agreement with 
the theoretical predictions.
First, we have performed two different sets of fits
in order to determine $D$ and $\delta$. As before, we consider
\bea
\log (f(\rho)) &=& \log f_{\infty} - D \rho^\delta, 
\label{fit5.19} \\
\log (\rho^{-0.255}\ f(\rho)) &=& \log f_{\infty} - D \rho^\delta, 
\label{fit5.20}
\eea
for various values of $\rho_{\rm min}$. The second fit keeps into account
the presence of $\rho^\sigma$ using the theoretical prediction for 
$\sigma$, Eq. (\ref{sigma-d3}). The results of the fits 
for fixed values of $N$ and different $N_{sh}$ are reported in Table
\ref{larger-fit1}. No significant dependence on $N_{\rm min}$ and 
$\rho_{\rm min}$ is visible in these results. Considering all data with 
$N\ge 1000$, and using for each $N$ the largest $N_{sh}$ appearing 
in Table \ref{larger-fit1}, 
we obtain from the first fit
\bea
\delta &=& \cases{2.504 \pm 0.002 & \hskip 1truecm $\rho_{\rm min} = 3.0$, \cr
                  2.481 \pm 0.004 & \hskip 1truecm $\rho_{\rm min} = 3.5$,} 
\\ 
D &=& \cases{0.1222 \pm 0.0004 & \hskip 1truecm $\rho_{\rm min} = 3.0$, \cr
             0.1277 \pm 0.0008 & \hskip 1truecm $\rho_{\rm min} = 3.5$,}
\eea
while the second one gives
\bea
\delta &=& \cases{2.444 \pm 0.002 & \hskip 1truecm $\rho_{\rm min} = 3.0$, \cr
                  2.441 \pm 0.004 & \hskip 1truecm $\rho_{\rm min} = 3.5$,} 
\\ 
D &=& \cases{0.1397 \pm 0.0005 & \hskip 1truecm $\rho_{\rm min} = 3.0$, \cr
             0.1406 \pm 0.0009 & \hskip 1truecm $\rho_{\rm min} = 3.5$.}
\eea
Fit \reff{fit5.19} gives estimates 
that show strong scaling corrections, clearly due to the neglected 
power term. The asymptotic values are difficult to estimate from this 
fit. In any case, we should observe that the estimates have the 
correct trend towards the expected results.
The second fit is more stable. The estimates are in much better 
agreement with the theoretical results, although larger 
values of $\rho_{\rm min}$ are necessary to confirm the theory at
the level of the statistical precision we have here.
{}From the second fit we can also estimate $f_\infty$. We obtain
\beq
f_{\infty} = \cases{
       0.0158 \pm 0.0001 & \hskip 1truecm $\rho_{\rm min} = 3.0$, \cr
       0.0159 \pm 0.0001 & \hskip 1truecm $\rho_{\rm min} = 3.5$,}
\eeq
in agreement with Eq. (\ref{f-d3}). 

Finally, in order to obtain estimates of $f_\infty$ and $\sigma$,
we have performed a fit of the form
\beq
\log\left[ f(\rho) \exp(D \rho^\delta)\right] = 
\log f_{\infty} + \sigma \log \rho,
\eeq
using the theoretical estimates for $D$ and $\delta$, for various 
values of $\rho\ge \rho_{\rm min}$ (see Table \ref{larger-fit2}). 
These fits become rapidly unstable with increasing $\rho_{\rm min}$. 
Nonetheless, as we shall see, the final results are in reasonable agreement with
the theoretical estimates even if one considers $1 \le \rho_{\rm min} \le 2$.
Fitting all data with $N\ge 1000$, we obtain
\bea
\sigma &=& \cases{
     0.2454 \pm 0.0003 & \hskip 1truecm $\rho_{\rm min} = 1.0$, \cr
     0.2426 \pm 0.0004 & \hskip 1truecm $\rho_{\rm min} = 1.5$, \cr
     0.2351 \pm 0.0006 & \hskip 1truecm $\rho_{\rm min} = 2.0$, }
\\
f_{\infty} &=& \cases{
      0.01608 \pm 0.00001 & \hskip 1truecm $\rho_{\rm min} = 1.0$, \cr
      0.01612 \pm 0.00001 & \hskip 1truecm $\rho_{\rm min} = 1.5$, \cr
      0.01625 \pm 0.00002 & \hskip 1truecm $\rho_{\rm min} = 2.0$. }
\eea
These results are not far from  the theoretical estimates, although they show 
a trend with increasing $\rho_{\rm min}$ which is the opposite of what
one expects theoretically: indeed the difference between the 
numerical and the theoretical estimates increases with 
$\rho_{\rm min}\to\infty$. It should be noticed however that this 
behaviour could be a result of corrections to scaling: 
indeed, for $\rho_{\rm min} = 2$, one observes that the estimates of 
$\sigma$ increase with $N$. 
If we analyze together all data with $N\ge N_{\rm min}$, 
we have ($\rho_{\rm min} = 2$)
\bea
\sigma &=& \cases{  
    0.2351 \pm 0.0006 & \hskip 1truecm $N_{\rm min} = 1000$, \cr
    0.2367 \pm 0.0006 & \hskip 1truecm $N_{\rm min} = 2000$, \cr
    0.2379 \pm 0.0007 & \hskip 1truecm $N_{\rm min} = 4000$, \cr
    0.2393 \pm 0.0009 & \hskip 1truecm $N_{\rm min} = 8000$, }
\\
f_{\infty} &=& \cases{
    0.01625 \pm 0.00001 & \hskip 1truecm $N_{\rm min} = 1000$, \cr
    0.01622 \pm 0.00001 & \hskip 1truecm $N_{\rm min} = 2000$, \cr
    0.01619 \pm 0.00001 & \hskip 1truecm $N_{\rm min} = 4000$, \cr
    0.01616 \pm 0.00002 & \hskip 1truecm $N_{\rm min} = 8000$. }
\eea
Therefore, we observe two opposite effects: $\sigma$ increases with 
increasing $\rho_{\rm min}$ because of non-asymptotic 
corrections in $f(\rho)$, while it decreases with increasing 
$N_{\rm min}$ because of corrections to scaling. Such a behaviour is 
not unexpected, since $f_w(\rho)$ was found to behave in 
exactly the same manner. For these reasons an accurate numerical check of
the predictions for $\sigma$ and $f_\infty$ is difficult: it is however
reassuring that all estimates are reasonably near the theoretical results.

Finally, we have studied the behaviour of $f(\rho)$ for $\rho \to 0$.
In this case we have performed fits of the form
\beq
\log f(\rho) = \log f_1 + \theta \log \rho,
\eeq
using, in each case, only data with 
$\rho_{\rm min} < \rho < \rho_{\rm max}$. 
We have introduced here two cuts, $\rho_{\rm min}$ and 
$\rho_{\rm max}$. The meaning of the latter is clear: 
it plays the role of $\rho_{\rm min}$ in the analysis of the 
large-$\rho$ behaviour of the EEDF. The second parameter $\rho_{\rm min}$
is introduced to eliminate spurious lattice effects. Indeed, the 
scaling limit is obtained taking $|\bf r|$ to infinity. In other words,
small values of $r$ should be discarded. In our fits we have taken 
$\rho_{\rm min} = 0.1$. Of course, this choice introduces a bias, 
and one should study the limit $\rho_{\rm min}\to 0$ to obtain 
the correct asymptotic behaviour. In our case, the systematic error
appears to be small: indeed the estimates are stable 
with respect to small changes of this parameter.
The results for fixed values of $N$ 
are reported in Table \ref{smallr}. The data 
show a small systematic variation with $N$.
Using all data with $N \ge 2000$, we obtain
\bea
\theta &=& \cases{
     0.223 \pm 0.003 & \hskip 1truecm $\rho_{\rm max} = 0.6$, \cr
     0.253 \pm 0.006 & \hskip 1truecm $\rho_{\rm max} = 0.4$, \cr
     0.281 \pm 0.031 & \hskip 1truecm $\rho_{\rm max} = 0.2$, }
\\
f_1 &=& \cases{
      0.01539 \pm 0.00004 & \hskip 1truecm $\rho_{\rm max} = 0.6$, \cr
      0.01599 \pm 0.00012 & \hskip 1truecm $\rho_{\rm max} = 0.4$, \cr
      0.01680 \pm 0.00096 & \hskip 1truecm $\rho_{\rm max} = 0.2$. }
\eea
The estimates of $\theta$ show a systematic variation with
$\rho_{\rm max}$, indicating the presence of strong non-asymptotic 
corrections. For $\rho_{\rm max} \ltapprox 0.40$, $\theta$
is in reasonable agreement with the theoretical prediction. 
The constant $f_1$ shows a similar trend, approaching the theoretical value 
for $\rho_{\rm max} \ltapprox 0.40$.
{}From $f_1$ we can compute $f_0 = f_1 2^\theta$. Using the theoretical 
prediction for $\theta$, we have $f_0 = 0.01853(5),
0.01925(14),0.0202(11)$, corresponding respectively to 
$\rho_{\rm max} = 0.6$, $0.4$, $0.2$. These estimates are in reasonable
agreement with the theoretical results
presented before that predicted $0.015\ltapprox f_0 \ltapprox 0.019$.

In conclusion, our Monte Carlo results confirm
the theoretical results of the previous Sections. 
Notice that the theoretical predictions are more precise than
the Monte Carlo estimates, in spite of the large statistics and of 
the very long walks used.

\section*{Acknowledgements}

We thank Peter Grassberger for useful comments on the first 
draft of this work.

\appendix
\section{Definitions and properties of the basic functions}

In this Appendix we report the definitions and asymptotic 
expansions of the functions that appear in our two-loop
computation of the end-to-end distribution function.

\bigskip

\noindent
{\em Function $N_1(x)$}

\bigskip

We define 
\be
N_1(x) = \, \int_0^\infty dt\, (1 - x t)\ e^{-x t}\ \log t
          \ \log |1 - t|. 
\ee
The expansion of $N_1(x)$ for large values of $x$ is easily computed.
Indeed, in this limit the relevant contribution is due to the region
$t\approx 0$. It is then enough to expand $\log |1 - t|$ in powers of
$t$ and integrate term by term. One obtains
\be
N_1(x) = \, \sum_{n=1}^\infty {(n-1)!\over x^{n+1}}\left[n \psi(n+1) + 1\right]
   -\, \log x\ \sum_{n=1}^\infty {n!\over x^{n+1}},
\ee
where
$\psi(x)$ is the logarithmic derivative of Euler's $\Gamma$-function. 
To obtain the asymptotic expansion for small values of $x$, one first
notices that $N_1(x)$ satisfies the differential equation
\be
N_1'(x) + N_1(x) =\, {1\over x^2}\left[2 - \gamma_E - \log x - 
    e^{-x} {\rm Ei}(x)\right],
\ee
where 
$\gamma_E\approx 0.5772156649$ is Euler's constant, and ${\rm Ei}$ is 
the exponential integral function \cite{Gradshtein}. 
Solving the previous equation, one obtains 
a different integral representation for $N_1(x)$:
\bea
\hskip -1truecm
&& N_1(x) = e^{-x} \int_0^x {dt\over t^2} 
   \left[e^t\left(2 - \gamma_E - \log t\right) - {\rm Ei}(t) 
        - (2 + t) (1 - \gamma_E - \log t)\right] 
\nonumber \\
\hskip -1truecm
&& \quad
+ e^{-x}\left[ {2\over x} (\gamma_E + \log x) +
                  (1 - \gamma_E) \log x - \smfrac{1}{2} \log^2 x + 1
          + \gamma_E - \smfrac{5}{12} \pi^2 - \smfrac{1}{2} \gamma_E^2 \right].
\eea
Using this expression it is trivial to obtain the expansion of $N_1(x)$ for 
$x\to 0$:
\bea
\hskip -1truecm 
N_1(x) &=& {2\over x} (\log x + \gamma_E) + 
   \smfrac{1}{2}\left[2 - 2 \gamma_E - \gamma_E^2 - \smfrac{5}{6}\pi^2
      - 2 (1 + \gamma_E) \log x - \log^2 x\right] 
\nonumber \\
\hskip -1truecm 
&& + {x\over2}\left[\smfrac{1}{2} - \gamma_E + \gamma_E^2 + \smfrac{5}{6}\pi^2
       + (2 \gamma_E - 1)\log x + \log^2 x\right] + O(x^2 \log x).
\eea
\bigskip

\noindent
{\em Function $N_2(x)$}

\bigskip

We define
\be
N_2(x) = \, \int_0^\infty {dt\over t} 
    J_1(t) 
    \int_C {dz\over 2\pi i} {z^{3/2} e^{xz}\over (z+1)^2 }
       \left[ K_1^3\left({\sqrt{z} t}\right) - 
              H_r\left({\sqrt{z} t}\right)\right],
\ee
where $C$ is a loop contour going counterclockwise around the 
negative real axis, $K_1(x)$ and $J_1(x)$ are Bessel functions
\cite{Gradshtein}, and 
$H_r(t)$ is defined by
\bea
H_r(t) &=& e^{-rt}\left[{1\over t^3} + {r\over t^2} + 
      {1\over 2t}\left(-\smfrac{3}{2} + r^2 + 3 \gamma_E + 
            3 \log {t\over2}\right)\right.
\nonumber \\ 
&& \left. \qquad - {3r\over4} + {r^3\over 6} + 
            {3 r\over 2}\gamma_E + {3r\over2}\log {t\over2}\right].
\label{defHr}
\eea
Note that $H_r(t)$ is such that 
$K^3_1(t)-H_r(t)\sim t\log^2 t$ for $|t|\to 0$.

We want now to derive the asymptotic behaviour of $N_2(x)$ for large
and small values of $x$. 
Substituting $w=x z$ and $s = t/\sqrt{x}$ we can rewrite 
\be
N_2(x) = \, x^{-1/2}\int_0^\infty {ds\over s} 
    J_1(s \sqrt{x})
    \int_C {dw\over 2\pi i} {w^{3/2} e^{w}\over (w+x)^2 }
       \left[K_1^3\left({\sqrt{w} s}\right) - 
             H_r\left({\sqrt{w} s}\right)\right].
\ee
To derive the small-$x$ behaviour, we expand $J_1(s \sqrt{x})$ 
(the corresponding series converges everywhere since $J_1(z)$ is 
an entire function) obtaining
\be
N_2(x) = \, \sum_{n=0}^\infty {(-x)^n\over n! (n+1)!}
   \int_0^\infty {ds\over s} \left({s\over2}\right)^{2n+1}
       \int_C {dw\over 2\pi i} {w^{3/2} e^{w}\over (w+x)^2 }
       \left[K_1^3\left({\sqrt{w} s}\right) - 
           H_r\left({\sqrt{w} s}\right)\right].
\ee
Now, $K_1(z) \sim z^{-1/2} e^{-z}$ for $|z|\to\infty$. Since 
$|{\rm arg}\, w|<\pi$, we can define $t=\sqrt{w} s$, and rotate the 
contour so that $t$ belongs to the positive real axis. We obtain
\be
N_2(x) = \, \sum_{n=0}^\infty {(-x)^n\over n! (n+1)!}
   \int_0^\infty {dt\over t} \left({t\over2}\right)^{2n+1} 
        \left[K_1^3(t) - H_r(t)\right]
       \int_C {dw\over 2\pi i} {w^{1-n} e^{w}\over (w+x)^2 }.
\label{eqA.13}
\ee
The last integral can be done exactly. We get
\be
\int_C {dw\over 2\pi i} {w^{1-n} e^{w}\over (w+x)^2 } = 
 (-1)^{n+1} x^{-n} e^{-x} (x + n -1) + 
 \sum_{k=0}^{n-2} {n-k-1\over k!} (-x)^{k-n},
\label{eq.A15}
\ee
with the convention that, for $n<2$, the summation is zero.
Substituting in Eq. (\ref{eqA.13}) we obtain
\bea
N_2(x) &=& 
  - e^{-x} \int_0^\infty {dt\over t} [K_1^3(t) - H_r(t)]
  \left[ {t\over2} I_0(t) + (x-2) I_1(t) \right]
\nonumber \\
&&  
 + \sum_{k=0}^\infty {(-x)^{k}\over k!} \,
   \int_0^\infty {dt\over t} [K^3_1(t) - H_r(t)]
\nonumber \\
&& \qquad \times
    \left[ {t\over2} I_0(t) - (k+2) I_1(t) -
    \sum_{n=0}^k {n-k-1\over n! (n+1)!} \left({t\over2}\right)^{2n+1}
    \right].
\label{N2-smallx-1}
\eea
This expression can be simplified and one obtains the 
final result
\be
N_2(x)=\, 
   \sum_{k=0}^\infty {(-x)^{k}\over k!} \,
   \sum_{n=0}^k {k+1-n\over n! (n+1)!}
   \int_0^\infty {dt\over t} [K^3_1(t) - H_r(t)]
   \left({t\over2}\right)^{2n+1}.
\ee
This expansion converges absolutely for all values of $x$ and allows
the computation of $N_2(x)$ up to quite large values of $x$ with a small 
effort. It gives immediately 
the small-$x$ expansion of $N_2(x)$. 

To compute the large-$x$ behaviour of $N_2(x)$,
we start by introducing the Mellin transforms of $[K^3_1(t) - H_r(t)]/t$ 
and of $J_1(t)$. Explicitly, we define
\be
M(z) \equiv \, \int_0^\infty dt\, t^{z-2}\left[ K_1^3(t) - H_r(t)\right],
\label{defMz}
\ee
and compute 
\be
\int_0^\infty dt\, t^{z-1} J_1(t) =\, 
   2^{z-1} {\Gamma({z+1\over2})\over \Gamma({3-z\over2})}.
\label{Mellintr-J1}
\ee
Eqs. \reff{defMz} and \reff{Mellintr-J1} are defined for ${\rm Re}\, z>0$
and $-1 < {\rm Re}\, z< {3\over2}$ respectively. Using
the Parseval formula for Mellin transforms, 
we can rewrite $N_2(x)$ as 
\be
N_2(x) = x^{-1/2} \int_C {dw\over 2\pi i} {w^2 e^w\over (w + x)^2}
 \int_{\gamma-i \infty}^{\gamma + i \infty} {dz\over 2\pi i} 
 M(1-z) x^{-z/2} (4 w)^{(z-1)/2} 
 {\Gamma({z+1\over2})\over \Gamma({3-z\over2})},
\ee
where $\gamma$ is any real number with $-1<\gamma<1$. Let us now show that
$N_2(x)\to 0$ for $x\to\infty$ faster than $x^{-3+\epsilon}$, for any 
$\epsilon>0$. Indeed
\be
|N_2(x)| \le\, x^{-1/2 - \gamma/2} 2^{\gamma-1} \int_C {|dw|\over 2\pi } 
     {e^{{\rm Re}\, w} \over |w + x|^2} |w|^{(\gamma + 3)/2} f(w),
\label{bound-on-N2}
\ee
where
\be
f(w)\equiv\, \int_{- \infty}^{\infty} {dy\over 2\pi}
   |M(1 - \gamma - i y)| e^{-y ({\rm arg}\, w)/2}
   \left| 
   {\Gamma({\gamma + iy +1\over2})\over \Gamma({3-\gamma - i y\over2})}\right|.
\ee
Let us first consider the integral $f(w)$.
For $|y|\to \infty$, an easy computation shows that 
$|M(1 - \gamma - i y)| \sim |y|^{p(\gamma)} e^{- \pi y/2}$, where 
$p(\gamma)$ is an exponent we do not need to know explicitly. 
In the same limit the ratio of $\Gamma$-functions behaves as 
$|y|^{\gamma - 1}$. Thus, the integral  $f(w)$ is finite for 
$|{\rm arg}\, w| < \pi$. For $|{\rm arg}\, w| \to \pi$, depending on the 
value of $\gamma$, the integral may be finite or diverge as a power
of $(|{\rm arg}\, w| - \pi)$. But, with our choice of $C$, 
$|{\rm arg}\, w| \to \pi$ corresponds to ${\rm Re}\, w \to -\infty$. 
In this limit the integrand in \reff{bound-on-N2} decreases exponentially, 
therefore making the whole expression finite. 
Finally notice that
\be
{1\over |w + x|^2} \le {4\over x^2}\left[1 - \chi\left(-{3x\over2}\le 
   {\rm Re}\, w\le -{x\over2}\right)\right] + 
   {1\over ({\rm Im}\, w)^2} \chi\left(-{3x\over2}\le 
   {\rm Re}\, w\le -{x\over2}\right),
\ee
where $\chi({\rm condition})$ is 1 if the condition is satisfied, 0
otherwise. It follows
\bea
|N_2(x)| &\le& x^{-5/2 - \gamma/2} 2^{\gamma+1} \int_C {|dw|\over 2\pi }
     {e^{{\rm Re}\, w}} |w|^{(\gamma + 3)/2} f(w)
\nonumber \\
&& + x^{-1/2 - \gamma/2} 2^{\gamma-1} \int_{\overline{C}} {|dw|\over 2\pi }
     {e^{{\rm Re}\, w}\over ({\rm Im}\, w)^2} |w|^{(\gamma + 3)/2} f(w),
\label{N2-bound-2}
\eea
where ${\overline{C}}$ is the part of $C$ with $-{3x\over2}\le
   {\rm Re}\, w\le -{x\over2}$. It is possible to choose $C$ so that 
$|{\rm Im}\, w|$ is constant along ${\overline{C}}$. It is then trivial 
to show that the second integral in Eq. \reff{N2-bound-2} is bounded
by $x^p e^{-x/2}$, where $p$ is an appropriate power.
Thus 
$|N_2(x)|< {\rm const}\ x^{-5/2 - \gamma/2}$. Since $\gamma$ is arbitrary
with $\gamma < 1$, the result follows immediately. A little more 
work, using the technique presented for $N_3(x)$, allows to show that
$N_2(x)\sim O(x^{-3} \log^2 x)$.
It is interesting to 
notice that, by adding additional terms in $H_r(x)$, one can make 
$N_2(x)$ decrease faster: if $H_r(x)$ is such that 
$K_1^3(x) - H_r(x) \sim O(x^{n} \log^3 x)$, then $N_2(x)$ decreases 
faster than $x^{-n/2-2+\epsilon}$, for any $\epsilon > 0$ (more precisely
$N_2(x)\sim O(x^{-n/2-2}\log^3 x)$).

\bigskip

\noindent
{\em Function $N_3(x)$}

\bigskip

We define
\be
N_3(x) = \, \int_0^\infty {dt\over t} 
    \left[J_1(t) - \smfrac{1}{2} t + \smfrac{1}{16} t^3\right]
    \int_C {dz\over 2\pi i} {z^{3/2} e^{xz}\over (z+1)^2 }
       H_r\left({\sqrt{z} t}\right),
\ee
where $C$ is a loop contour going counterclockwise around the 
negative real axis, $J_1(x)$ is a Bessel function \cite{Gradshtein}
and $H_r(t)$ is defined in Eq. (\ref{defHr}).

The small-$x$ behaviour is easily computed using the same procedure 
as before. We obtain
\be
N_3(x) = \,
  \sum_{k=2}^\infty {(-x)^{k}\over k!} \sum_{n=2}^k 
   {k+1-n\over n! (n+1)!} 2^{-2n - 1} \widehat{H}_r(2n+2),
\ee
where $\widehat{H}_r(z)$ is the Mellin transform of $H_r(t)/t$:
\bea
\widehat{H}_r(z)&\equiv&
   \int_0^\infty dt\, t^{z-2} H_r(t) 
\nonumber \\
&=& r^{2-z} \left[ r^2 \Gamma(z-4) + r^2 \Gamma(z-3) + 
   \smfrac{1}{4} \Gamma(z-2)(-3 + 6\gamma_E + 2 r^2) \right.
\nonumber \\
&& \left. + \smfrac{1}{12} \Gamma(z-1) (-9 + 18 \gamma_E  + 2 r^2)+
      \smfrac{3}{2} \Gamma(z-2) (\psi(z-2) - \log 2r) \right.
\nonumber \\
&& \left. \qquad + 
      \smfrac{3}{2} \Gamma(z-1) (\psi(z-1) - \log 2r)\right].
\eea
This expansion converges absolutely for all values of $x$ and 
it gives immediately 
the small-$x$ expansion of $N_3(x)$. 

Let us now compute the behaviour for large values of $x$. 
We rewrite $N_3(x)$ as 
\be
N_3(x) = L_1(x) + L_2(x) ,
\ee
where
\bea
L_1(x) &=& x^{-1/2} \int_0^\infty {ds\over s}
   \left[J_1(\sqrt{x} s) + \theta(1 - \sqrt{x} s)
   \left(- \smfrac{1}{2} s x^{1/2} +
    \smfrac{1}{16} s^3 x^{3/2}\right)\right]
\nonumber \\
&& \quad 
  \times \int_C {dw\over 2\pi i} {w^{3/2} e^w\over (w + x)^2} H_r(\sqrt{w} s),
\\
L_2(x) &=& x^{-1/2} \int_0^\infty {ds\over s}
   \theta(\sqrt{x} s - 1)
   \left(- \smfrac{1}{2} s x^{1/2} +
    \smfrac{1}{16} s^3 x^{3/2}\right)
\nonumber \\
&& \quad 
  \times \int_C {dw\over 2\pi i} {w^{3/2} e^w\over (w + x)^2} H_r(\sqrt{w} s),
\eea
and $\theta(x)$ is Heaviside's step function.

Let us first consider $L_1(x)$. Using the Parseval formula for 
Mellin transforms we obtain 
\be 
L_1(x) =\, x^{-1/2} 
\int_C {dw\over 2\pi i} {w^{3/2} e^w\over (w + x)^2}
\int_{\gamma - i \infty}^{\gamma + i \infty} 
 {dz\over 2\pi i} 
 w^{z/2} x^{-z/2} \widehat{H}(1-z) B(z),
\ee
where
\be
B(z) =\,
 2^{z-1} {\Gamma({z+1\over2})\over \Gamma({3-z\over2})} -
   \smfrac{1}{2} {1\over z+1} + \smfrac{1}{16} {1\over z + 3},
\ee
and $\gamma$ is a real number satisfying $-5<\gamma <-3$. 
Now, rewrite the previous expression for $L_1$ as 
\bea
L_1(x) &=& x^{-1/2}
\int_C {dw\over 2\pi i} {w^{3/2} e^w\over (w + x)^2}
\int_{\gamma - i \infty}^{\gamma + i \infty}
 {dz\over 2\pi i}
 w^{z/2} x^{-z/2} \widehat{H}(1-z) B(z)
\nonumber \\
&& - x^{-1/2}
\int_C {dw\over 2\pi i} {w^{3/2} e^w\over (w + x)^2}
 \sum_{n=-3}^0 \mathop{\rm Res}_{z=n}
  \left[w^{z/2} x^{-z/2} \widehat{H}(1-z) B(z)\right],
\label{L1-residui-integrale}
\eea
where, in the first integral,  $\gamma$ is real such that 
$0 < \gamma < 1$. Repeating the discussion presented for $N_2(x)$ one can 
easily show that the first integral in Eq. \reff{L1-residui-integrale}
behaves as $O(x^{-3+\epsilon})$. Neglecting terms of this order, we obtain
\bea
L_1(x) &=& {x\over64} (- 4 \log 2 + 4 \gamma_E + 11) 
  \int_C {dw\over 2\pi i} {e^w\over (w + x)^2} 
\nonumber \\
&& - \left ({3\over64} - {27\over 64} \gamma_E + {3\over8} \gamma_E^2
     + {27\over 64} \log 2 + {3\over8} \log^2 2 - 
       {3\over 4} \gamma_E  \log 2\right)
  \int_C {dw\over 2\pi i} {w e^w\over (w + x)^2} 
\nonumber \\
&& - \left( - {27\over 128} + {3\over 8}\gamma_E - {3\over 8} \log 2\right)
  \int_C {dw\over 2\pi i} {w e^w\over (w + x)^2}
  \log (w/x).
\eea
Using Eq. \reff{eq.A15} and $(n\ge 0)$
\be
\int_C {dw\over 2\pi i} {w^n e^w\over (w + x)^2} \log w =\, 
    {(-1)^{n+1} n!\over x^2}\left(1 + {2(n+1)\over x}\right) + O(x^{-4}),
\ee
we obtain
\be
L_1(x) = \left( {27\over 128} - {3\over 8}\gamma_E + {3\over 8} \log 2\right)
      {1\over x^2} + O(x^{-3} \log x).
\ee
The computation of the asymptotic behaviour of $L_2(x)$ is completely
analogous. We have, discarding terms of order $x^{-3}\log^2 x$,
\be
L_2(x) =\, - x^{-1/2} 
   \int_C {dw\over 2\pi i} {w^{3/2} e^w\over (w + x)^2}
 \sum_{n=-3}^0 \mathop{\rm Res}_{z=n}
  \left[w^{z/2} x^{-z/2} \widehat{H}(1-z) 
   \left( {1\over 2}{1\over z+1} - {1\over 16} {1\over z+3}\right) \right].
\ee
Using the previous results and $(n\ge 0)$
\be
\int_C {dw\over 2\pi i} {w^n e^w\over (w + x)^2} \log^2 w =\, 
    {2 (-1)^{n+1} \psi(n+1) n!\over x^2} + O(x^{-3}),
\ee
we obtain 
\be
L_2(x) =\, {1\over 32 x^2} \left[ x + {5\over4} + 6 \gamma_E - 
   12 \log 2 - 6 \log x\right] + O(x^{-3} \log^2 x).
\ee
Summing up, we have 
\be
N_3(x) =\, {1\over 32 x^2} \left[ x + 8 - 6 \gamma_E - 6 \log x\right] +
             O(x^{-3} \log^2 x).
\ee

\bigskip

\noindent
{\em Fourier transform of $\widehat{N}_2(Q^2) + N_3(Q^2)$}

\bigskip

We want now to report the small-$\rho$ behaviour of the 
Fourier transform of $\widehat{N}_2(Q^2) + N_3(Q^2)$ that 
we use in Sec. \ref{sec3}. We define
\be
N(\rho) \equiv \, 
  \int {d^4Q\over (2 \pi)^4} \, e^{i\bf Q\cdot \rho}
  \left(\widehat{N}_2(Q^2) + N_3(Q^2)\right),
\ee
where, cf. Eq. \reff{definizione-hat},
\be
\widehat{N}_2(x) = N_2(x) - e^{-x} (1 + x) N_2(0) - 
    e^{-x} {N'}_2(0).
\ee
Using the definitions of $N_2(x)$ and $N_3(x)$ we can rewrite
\be
\widehat{N}_2(x) + N_3(x) = 
  x^{-1/2} \int_C {dw\over 2\pi i}
   {w^{3/2} e^w\over (w + x)^2}
  \int_0^\infty {ds\over s} K_1^3(s\sqrt{w})
  \left[J_1(\sqrt{x}s) - \smfrac{1}{2} \sqrt{x}s + 
            \smfrac{1}{16} x^{3/2} s^3\right].
\ee
Using the fact that, for any function $h(Q^2)$, we have
\be
\int {d^4Q\over (2 \pi)^4} \, e^{i\bf Q\cdot \rho} h(Q^2) =\, 
{1\over 4\pi^2 \rho} \int_0^\infty dQ\, Q^2 J_1(Q\rho) h(Q^2),
\ee
we can perform the integral over $Q$, cf. formula 6.541 of 
Ref. \cite{Gradshtein}, obtaining
\bea
N(\rho) &=& 
  {1\over 16\pi^2 \rho} \int_C {dw\over 2\pi i} w e^w\, 
   \left\{ \int_0^\infty {ds\over s} K_1^3(s)
   \left[ - {s\over \sqrt{w}} K_1(\sqrt{w}\rho)
   \left( I_2(s) + I_0(s) - 1 - \smfrac{3}{8} s^2\right) \right. \right.
\nonumber \\
 && \left. \left. 
    + \left(K_2(\sqrt{w}\rho) + K_0(\sqrt{w}\rho)\right)
      \left(I_1(s) - \smfrac{1}{2} s - \smfrac{1}{16} s^3\right)
    \vphantom{{ds\over s}}\right]\right\}
\nonumber \\
 && 
  + {1\over 16\pi^2 \rho} \int_C {dw\over 2\pi i} w e^w\,
    \left\{-\rho [I_0(\rho\sqrt{w}) + I_2(\rho\sqrt{w})]
    \int_\rho^\infty {ds\over s} K_1^4(s\sqrt{w}) \right.
\nonumber \\
 && \qquad
   - \rho [K_0(\rho\sqrt{w}) + K_2(\rho\sqrt{w})]
    \int_\rho^\infty {ds\over s} K_1^3(s\sqrt{w})I_1(s\sqrt{w})
\nonumber \\
 && \qquad
   + K_1(\rho\sqrt{w})
    \int_\rho^\infty {ds} K_1^3(s\sqrt{w})
        [I_2(s\sqrt{w}) + I_0(s\sqrt{w})]
\nonumber \\
 && \qquad \left. + I_1(\rho\sqrt{w})
    \int_\rho^\infty {ds} K_1^3(s\sqrt{w})
        [K_2(s\sqrt{w}) + K_0(s\sqrt{w})] \right\}.
\eea
Expanding for $\rho\to0$, after a lengthy calculation, we obtain
\bea
N(\rho) &\approx& {1\over 128\pi^2 \rho^2} - 
    {3\over 128\pi^2} \log^2{\rho\over2} - 
     {1\over 128\pi^2} (1 + 3 \gamma_E) \log{\rho\over2} 
\nonumber \\
&& + {5\over 1024} - {3\over 256\pi^2} - {\gamma_E\over 256\pi^2}
- {3\over 512\pi^2} \gamma_E^2 + 
 {A\over 32\pi^2} + O(\rho^2\log^2 \rho),
\eea
where
\be
A =\, 
  \int_0^\infty ds\left[
   K_1^3(s)\left(1 + \smfrac{1}{4}s^2\right) - 
   e^{-s/2}\left({1\over s^3} + {1\over 2s^2} + 
         {3\over 2s}\gamma_E - {3\over 8s} + {3\over 2s}\log{s\over2}\right)
   \right]\; .
\label{definizioneintA}
\ee
Numerically $A\approx - 1.30204$.

\bigskip

\noindent
{\em Function $g_1(x)$}

\bigskip

We define
\be
g_1(x) = \smfrac{1}{2} (1 - e^{-x}) + 
         \smfrac{1}{2} (1-x) e^{-x}\left({\rm Ei}(x) - \log x\right),
\ee
where ${\rm Ei}(x)$ is the exponential integral function \cite{Gradshtein}. 
We are interested in the behaviour of $g_1(x)$ for large and small 
values of $x$. These expansions are easily obtained using the corresponding
results for ${\rm Ei}(x)$. For $x\ll 1$ we obtain
\bea
g_1(x) &=& 
   - {1\over2} \sum_{n=0}^\infty {(n+1)\psi(n+1)\over n!} (-x)^n
\nonumber \\
&=&
 \smfrac{1}{2}\gamma_E + (1 - \gamma_E) x + 
    \smfrac{3}{8}(2\gamma_E - 3) x^2 + O(x^3).
\eea
For $x\gg 1$ we have the asymptotic expansion
\be
g_1(x) =\, - {1\over2}\ \sum_{n=0}^\infty
     {(n+1) (n+1)!\over x^{2+n}}.
\ee

\bigskip

\noindent
{\em Function $g_2(x)$}

\bigskip
    
We define
\bea
g_2(x) &=& {1\over 4x}\left[e^{-x} {\rm Ei}(x) + \gamma_E + \log x\right] -
        {1\over4} \log x 
\nonumber \\
&& + {1\over8} e^{-x} (1-x) \log x\left[\log x - 2 {\rm Ei}(x)\right]
\nonumber \\
&& + {1\over8} e^{-x}\left[2 \log x - 4 {\rm Ei}(x) - \pi^2 (1-x)\right]
   - {1\over 4} (1-x) N_1(x).
\eea
For $x\ll 1$ we have
\bea
g_2(x) &=& {1\over 8} \sum_{n=0}^\infty (n+1) 
    {\psi^2(n+1) - \psi'(n+1)\over n!} (-x)^n
\nonumber \\
&=& {1\over8} \left(\gamma_E^2 - \smfrac{1}{6}\pi^2\right) + 
    {x\over 4}\left(-2 + 2 \gamma_E - \gamma_E^2 + \smfrac{1}{6}\pi^2\right)
    + O(x^2).
\eea
For $x\gg 1$ we have
\be
g_2(x) =\, {1\over4} \sum_{k=1}^\infty
       {k! k\psi(k+1)\over x^{k+1}}.
\ee

\bigskip

\noindent
{\em Function $g_3(x)$}

\bigskip

We define
\be
g_3(x)=\, -1 + x e^{-x} {\rm Ei}(x) + e^{-x} (1 - x \log x).
\ee
For $x\ll 1$ we have
\bea
g_3(x) &=& \sum_{n=1}^\infty  
    {\psi(n+1)\over (n-1)!} (-x)^n
\nonumber \\
&=& 
    x\left(\gamma_E - 1\right) + 
    {x^2\over2} (3 - 2\gamma_E) + 
    + O(x^3).
\eea
For $x\gg 1$ we obtain
\be
g_3(x) = \sum_{k=1}^\infty {k!\over x^k}.
\ee

\clearpage

\clearpage


\begin{table}
\footnotesize
\begin{center}
\begin{tabular}{|c|c|c|c|c|}
\hline
$N$   &  $M_4$   & $M_6$  & $M_8$ & $M_{10}$  \\ 
\hline
500  & $1.49998 \pm 0.00042$  & $2.9413 \pm 0.0024$  & 
       $7.044 \pm 0.011$ & $19.785 \pm 0.057$   \\
1000  & $1.50457 \pm 0.00035$  & $2.9666 \pm 0.0020$ & 
       $7.1600 \pm 0.0098$ & $20.306 \pm 0.050$   \\
2000  & $1.50755 \pm 0.00034$  & $2.9833 \pm 0.0019$ & 
       $7.2360 \pm 0.0096$ & $20.647 \pm 0.050$   \\
4000  & $1.50939 \pm 0.00035$  & $2.9933 \pm 0.0020$ & 
       $7.2811 \pm 0.0095$ & $20.849 \pm 0.050$   \\
8000  & $1.51084 \pm 0.00026$  & $3.0011 \pm 0.0015$ & 
       $7.3153 \pm 0.0075$ & $20.996 \pm 0.039$   \\
16000  & $1.51169 \pm 0.00043$ & $3.0053 \pm 0.0026$ & 
       $7.334 \pm 0.012$ & $21.091 \pm 0.066$   \\
32000  & $1.51340 \pm 0.00068$ & $3.0158 \pm 0.0042$ & 
       $7.386 \pm 0.023$ & $21.34 \pm 0.11$   \\ 
\hline
$M^*$    
 & $1.51406 \pm 0.00089$ & $3.018 \pm 0.005$ & 
   $7.387 \pm 0.023$ & $21.34 \pm 0.13$ \\
B    & $-(0.00314 \pm 0.00084)$  & $-(0.016 \pm 0.0046)$ & $-(0.070 \pm 0.021)$ & $-(0.32 \pm 0.11)$ \\
$\Delta$ & $0.538 \pm 0.079$  & $0.555 \pm 0.085$  & $0.572 \pm 0.092$  & $0.56 \pm 0.11$ \\
CL   &  70\%          & 66\% & 67\% & 66\% \\
\hline
th & $1.51397 \pm 0.00079$ & $3.018 \pm 0.004$ & $7.392 \pm 0.015$ & $21.35 \pm 0.06$\\
\hline
ph & $1.50876 \pm 0.00023$ & $2.993 \pm 0.001$ & $7.292 \pm 0.005$ & $20.94 \pm 0.02$\\
\hline
\end{tabular}
\end{center}
\caption{Monte Carlo results for the invariant ratios $M_{2k}$.
``CL" is the confidence level of the fit 
$M_{2k} = M_{2k}^* + B_{2k} (N/8000)^{-\Delta}$.
``th" is the theoretical prediction \reff{Mstar} and
``ph" the phenomenological prediction \reff{Mstarapprox}.
}
\label{Mratios}
\end{table}

\begin{table}
\footnotesize
\begin{center}
\begin{tabular}{|c|c|c|c|c|}
\hline
   &  $R_4$   & $R_6$  & $R_8$ & $R_{10}$  \\ 
\hline
500  & $0.60052 \pm 0.00038$  & $0.42958 \pm 0.00080$ & $0.3348 \pm 0.0013$ & $0.2748 \pm 0.0019$   \\
1000  & $0.60023 \pm 0.00032$  & $0.42896 \pm 0.00068$ & $0.3338 \pm 0.0011$ & $0.2731 \pm 0.0016$   \\
2000  & $0.60008 \pm 0.00030$  & $0.42872 \pm 0.00065$ & $0.3336 \pm 0.0010$ & $0.2732 \pm 0.0015$   \\
4000  & $0.59983 \pm 0.00031$  & $0.42811 \pm 0.00064$ & $0.3325 \pm 0.0010$ & $0.2713 \pm 0.0015$   \\
8000  & $0.60005 \pm 0.00024$  & $0.42866 \pm 0.00050$ & $0.33339 \pm 0.00080$ & $0.2727 \pm 0.0012$   \\
16000  & $0.59994 \pm 0.00038$  & $0.42833 \pm 0.00079$ & $0.3328 \pm 0.0013$ & $0.2720 \pm 0.0019$   \\
32000  & $0.60028 \pm 0.00065$  & $0.4295 \pm 0.0012$ & $0.3351 \pm 0.0024$ & $0.2756 \pm 0.0035$   \\ 
\hline
th  & ${3 \over 5} = 0.6$ & ${3 \over 7} = 0.42857142$ & 
    ${1 \over 3} = 0.33333333$ & ${3 \over 11} = 0.27272727$   \\
\hline
\end{tabular}
\end{center}
\caption{Monte Carlo results for the invariant ratio $R_{2k}$. 
``th" is the theoretical prediction.}
\label{Rratios}
\end{table}

\begin{table}
\protect\footnotesize
\begin{center}
\begin{tabular}{|c|c||c|c|c||c|c|c|}
\hline
$N$ &&
\multicolumn{3}{|c|}{$f_w(\rho)$}
&\multicolumn{3}{|c|}{$f_w(\rho) \cdot {\rho}^{0.169}$}\\
\hline

        &&$\rho_{\rm min} = 3$ & $\rho_{\rm min} = 3.5$ & $\rho_{\rm min} = 4$&
          $\rho_{\rm min} = 3$ & $\rho_{\rm min} = 3.5$ & $\rho_{\rm min} = 4$\\

\hline
500 & $\delta$ &
     2.489(16) &     2.480(38) &     2.54(10) &
     2.535(17) &     2.515(38) &     2.57(10) \\
& D &
     0.1328(38) &     0.1350(94) &     0.121(24) &
     0.1203(35) &     0.1246(87) &     0.113(22) \\
& $f_{w,\infty}$ &
     0.2395(55) &     0.244(17) &     0.216(49) &
     0.2618(58) &     0.271(18) &     0.245(54) \\
\hline
1000 & $\delta$ &
     2.434(13) &     2.388(30) &     2.415(73) &
     2.479(13) &     2.423(30) &     2.444(74) \\
& D &
     0.1453(33) &     0.1583(87) &     0.150(21) &
     0.1318(30) &     0.1462(81) &     0.141(20) \\
& $f_{w,\infty}$ &
     0.2520(47) &     0.276(16) &     0.260(45) &
     0.2747(50) &     0.305(17) &     0.293(50) \\
\hline
2000 & $\delta$ &
     2.420(13) &     2.434(29) &     2.476(84) &
     2.463(13) &     2.468(29) &     2.494(83) \\
& D &
     0.1483(34) &     0.1442(78) &     0.133(21) &
     0.1346(31) &     0.1336(72) &     0.127(20) \\
& $f_{w,\infty}$ &
     0.2506(48) &     0.244(13) &     0.224(43) &
     0.2735(50) &     0.272(14) &     0.259(48) \\
\hline
4000 & $\delta$ &
     2.407(13) &     2.447(29) &     2.456(74) &
     2.452(13) &     2.483(28) &     2.486(75) \\
& D &
     0.1517(33) &     0.1408(75) &     0.138(20) &
     0.1375(31) &     0.1297(68) &     0.129(19) \\
& $f_{w,\infty}$ &
     0.2541(47) &     0.235(12) &     0.232(40) &
     0.2767(50) &     0.261(13) &     0.262(44) \\
\hline
8000 & $\delta$ &
     2.4139(98) &     2.419(23) &     2.450(58) &
     2.4590(99) &     2.454(23) &     2.480(59) \\
& D &
     0.1490(25) &     0.1477(61) &     0.139(15) &
     0.1350(23) &     0.1362(57) &     0.129(15) \\
& $f_{w,\infty}$ &
     0.2464(35) &     0.244(10) &     0.224(30) &
     0.2684(37) &     0.271(11) &     0.253(33) \\
\hline
16000 & $\delta$ &
     2.370(21) &     2.423(51) &     2.36(15) &
     2.414(21) &     2.458(51) &     2.37(15) \\
& D &
     0.1617(61) &     0.147(14) &     0.167(48) &
     0.1465(55) &     0.136(13) &     0.160(45) \\
& $f_{w,\infty}$ &
     0.2658(86) &     0.241(22) &     0.29(10) &
     0.2883(88) &     0.268(24) &     0.33(11) \\
\hline
32000 & $\delta$ &
     2.433(32) &     2.382(83) &     1.69(22) &
     2.482(33) &     2.422(83) &     1.74(22) \\
& D &
     0.1435(79) &     0.157(24) &     0.62(28) &
     0.1292(73) &     0.144(22) &     0.56(25) \\
& $f_{w,\infty}$ &
     0.236(10) &     0.257(39) &     2.4 (2.3) &
     0.257(11) &     0.282(41) &     2.3 (2.1) \\
\hline
\end{tabular}
\end{center}
\caption{Results for the fit  $g(\rho)
 = f_{w,\infty} \exp(- D \rho^\delta)$
for $\rho > \rho_{\rm min}$. The first three columns refer
to $g(\rho) = f_{w,MC}(\rho)$, the last three columns to
$g(\rho) = f_{w,MC}(\rho) \rho^{0.169}$. Here
$f_{w,MC}(\rho)$ is the Monte Carlo ``wall-to-wall" EEDF.
}
\label{wall-fit1}
\end{table}

\begin{table}
\protect\footnotesize
\begin{center}
\begin{tabular}{|c|c||c|c|c|}
\hline
$N$ && $\rho_{\rm min} = 2. $ &$ \rho_{\rm min} = 2.5 $ &
       $\rho_{\rm min} = 3.$ \\
\hline
500 & $\sigma_w$ &
    $-$0.1153(31) &    $-$0.1752(63) &    $-$0.252(12) \\
& $f_{w,\infty}$&
    0.27406(81) &    0.2934(20) &    0.3232(50) \\
\hline
1000 & $\sigma_w$ &
    $-$0.12149(24) &    $-$0.1641(47) &    $-$0.234(10) \\
& $f_{w,\infty}$&
    0.27204(62) &    0.2855(15) &    0.3117(38) \\
\hline
2000 & $\sigma_w$ &
    $-$0.1254(24) &    $-$0.1527(45) &    $-$0.1902(97) \\
& $f_{w,\infty}$&
    0.27076(60) &    0.2792(14) &    0.2926(35) \\
\hline
4000 & $\sigma_w$ &
    $-$0.1268(22) &    $-$0.1513(44) &    $-$0.1954(94) \\
& $f_{w,\infty}$&
    0.26937(56) &    0.2770(13) &    0.2927(34) \\
\hline
8000 & $\sigma_w$ &
    $-$0.1280(17) &    $-$0.1417(34) &    $-$0.1537(72) \\
& $f_{w,\infty}$&
    0.26850(44) &    0.2727(10) &    0.2768(24) \\
\hline
16000 & $\sigma_w$ &
    $-$0.1314(34) &    $-$0.1473(70) &    $-$0.2001(17) \\
& $f_{w,\infty}$&
    0.26874(87) &    0.2735(21) &    0.2919(60) \\
\hline
32000 & $\sigma_w$ &
    $-$0.1087(51) &    $-$0.125(11) &    $-$0.090(33) \\
& $f_{w,\infty}$&
    0.2624(12) &    0.2670(33) &    0.256(10) \\
\hline
\end{tabular}
\end{center}
\caption{Results for the fit $f_{w,MC}(\rho) \exp(D \rho^\delta)
 = f_{w,\infty} \rho^{\sigma_w}$
for $\rho > \rho_{\rm min}$. $f_{w,MC}(\rho)$ is the Monte Carlo
``wall-to-wall" EEDF.
$D$ and $\delta$ have been set equal to the theoretical predictions,
Eqs. (\protect\ref{delta}) and (\protect\ref{Dconstant}). }
\label{wall-fit2}
\end{table}

\begin{table}
\footnotesize
\begin{center}
\begin{tabular}{|c|c|c||c|c|c||c|c|c|}
\hline
$N$ & $N_{sh}$ &&
\multicolumn{3}{|c|}{$f_{MC}(\rho)$} &
\multicolumn{3}{|c|}{$f_{MC}(\rho)\cdot {\rho}^{-0.255}$}\\
\hline
&&& $\rho_{\rm min} = 3$& $\rho_{\rm min} = 3.5$& $\rho_{\rm min} = 4$&
$\rho_{\rm min} = 3$& $\rho_{\rm min} = 3.5$& $\rho_{\rm min} = 4$\\
\hline
500 & 1 & $\delta$ &
   2.5368(59) &   2.479(11)  &   2.296(23)  &
   2.4733(58) &   2.430(11)  &   2.260(23)  \\
& & D &
   0.1170(12) &   0.1305(27)  &   0.1877(86)  &
   0.1346(14) &   0.1461(30)  &   0.2050(93)  \\
& & $f_\infty$ &
   0.01827(15)  &   0.02045(43)  &   0.0333(21)  &
   0.01583(14)  &   0.01732(38)  &   0.0279(18)  \\
& 10 & $\delta$ &
   2.5662(59) &   2.545(11)  &   2.464(23)  &
   2.5019(58) &   2.495(11)  &   2.425(23)  \\
& & D &
   0.1113(12) &   0.1158(24)  &   0.1359(61)  &
   0.1282(13) &   0.1299(26)  &   0.1490(66)  \\
& & $f_\infty$ &
   0.01762(14)  &   0.01833(36)  &   0.0224(12)  &
   0.01523(13)  &   0.01545(32)  &   0.0185(11)  \\
\hline
1000 & 1 & $\delta$ &
   2.4892(47) &   2.4078(88)  &   2.202(18)  &
   2.4268(46) &   2.3601(89)  &   2.167(18)  \\
& & D &
   0.1262(10) &   0.1473(24)  &   0.2216(81)  &
   0.1450(12) &   0.1646(27)  &   0.2419(88)  \\
& & $f_\infty$ &
   0.01905(13)  &   0.02244(39)  &   0.0395(21)  &
   0.01655(12)  &   0.01911(35)  &   0.0335(19)  \\
& 20 & $\delta$ &
   2.5336(47) &   2.5094(88)  &   2.461(19)  &
   2.4698(46) &   2.4596(87)  &   2.422(18)  \\
& & D &
   0.11696(96)  &   0.1224(20)  &   0.1345(51)  &
   0.1346(11) &   0.1373(21) &   0.1477(52)  \\
& & $f_\infty$ &
   0.01800(12)  &   0.01885(30)  &   0.02107(97)  &
   0.01560(11)  &   0.01593(27)  &   0.01744(79)  \\
\hline
2000 & 40 & $\delta$ &
   2.5123(44)  &   2.4940(83)  &   2.421(17)  &
   2.4489(43)  &   2.4442(81)  &   2.382(17)  \\
& & D &
   0.12072(94)  &   0.1249(19)  &   0.1443(48)  &
   0.1389(11)  &   0.1401(21) &   0.1585(53)  \\
& & $f_\infty$ &
   0.01822(12)  &   0.01885(29) &   0.02257(95)  &
   0.01580(10)  &   0.01595(25) &   0.01875(82)  \\
& 80 & $\delta$ &
   2.5161(44)  &   2.5035(83)  &   2.445(17)  &
   2.4526(43)  &   2.4534(83)  &   2.406(17)  \\
& & D &
   0.11993(94)  &   0.1228(19)  &   0.1376(46)  &
   0.1380(11)  &   0.1378(21)  &   0.1513(50)  \\
& & $f_\infty$ &
   0.01813(11) &   0.01856(28) &   0.02135(89)  &
   0.01573(10)  &   0.01569(25)  &   0.01771(76)  \\
\hline
4000 & 100 & $\delta$ &
   2.5031(42) &   2.4716(79) &   2.4289(16)  &
   2.4398(41) &   2.4221(79) &   2.390(17) \\
& & D &
   0.12229(91)  &   0.1297(19)  &   0.1413(45)  &
   0.1407(10) &   0.1455(22) &   0.1552(52)  \\
& & $f_\infty$ &
   0.01828(11)  &   0.01940(29)  &   0.02168(85)  &
   0.01587(10)  &   0.01644(26)  &   0.01799(78)  \\
& 200 & $\delta$ &
   2.5071(43) &   2.4823(79) &   2.457(16)  &
   2.4439(42) &   2.4325(78) &   2.418(17)  \\
& & D &
   0.12145(92)  &   0.1272(19)  &   0.1338(43)  &
   0.1397(10) &   0.1427(21) &   0.1471(49)  \\
& & $f_\infty$ &
   0.01818(11)  &   0.01905(28)  &   0.02036(79)  &
   0.01578(10)  &   0.01613(25)  &   0.01685(72)  \\
\hline
8000 & 200 & $\delta$ &
   2.4953(35) &   2.4839(67) &   2.447(14)  &
   2.4320(35) &   2.4341(65) &   2.408(14)  \\
& & D &
   0.12364(78) &   0.1263(16)  &   0.1358(36)  &
   0.14230(88) &   0.1418(17)  &   0.1493(42)  \\
& & $f_\infty$ &
   0.018323(94)  &   0.01872(23)  &   0.02048(67)  &
   0.015919(86)  &   0.01585(20)  &   0.01696(61)  \\
& 800 & $\delta$ &
   2.5003(36) &   2.4964(68) &   2.484(14)  &
   2.4369(35) &   2.4462(66) &   2.444(14)  \\
& & D &
   0.12258(78) &   0.1235(16)  &   0.1266(34)  &
   0.14108(88) &   0.1386(17)  &   0.1394(37)  \\
& & $f_\infty$ &
   0.018205(94)  &   0.01833(23)  &   0.01890(60)  &
   0.015808(87)  &   0.01551(20)  &   0.01561(52)  \\
\hline
16000 & 800 & $\delta$ &
   2.4913(53) &   2.4619(99) &   2.409(22)  &
   2.4283(52) &   2.4125(97) &   2.370(20)  \\
& & D &
   0.1245(12) &   0.1316(24) &   0.1461(62)  &
   0.1432(13) &   0.1475(27) &   0.1605(63)  \\
& & $f_\infty$ &
   0.01842(14)  &   0.01950(36)  &   0.0222(12)  &
   0.01601(13)  &   0.01654(32)  &   0.01847(96)  \\
& 1200 & $\delta$ &
   2.4935(53) &   2.466(10) &   2.423(22)  &
   2.4306(52) &   2.4169(98)  &   2.384(20)  \\
& & D &
   0.1240(12) &   0.1305(25)  &   0.1421(60)  &
   0.1426(13) &   0.1464(27)  &   0.1562(62)  \\
& & $f_\infty$ &
   0.01837(14)  &   0.01936(37)  &   0.0215(11)  &
   0.01595(13)  &   0.01641(32)  &   0.01786(92)  \\
\hline
32000 & 5000 & $\delta$ &
   2.4837(87)  &   2.479(16) &   2.366(33)  &
   2.4208(85)  &   2.429(16) &   2.327(32)  \\
& & D &
   0.1258(19)  &   0.1268(38) &   0.159(10)  &
   0.1446(22)  &   0.1422(42) &   0.174(11)  \\
& & $f_\infty$ &
   0.01844(23)  &   0.01855(55)  &   0.0246(21)  &
   0.01603(21)  &   0.01570(49)  &   0.0205(18)  \\
\hline
\end{tabular}
\end{center}
\caption{Results for the fit  $g(\rho)
 = f_{\infty} \exp(- D \rho^\delta)$
for $\rho > \rho_{\rm min}$. The first three columns refer
to $g(\rho) = f_{MC}(\rho)$, the last three columns to
$g(\rho) = f_{MC}(\rho) \rho^{-0.255}$. Here
$f_{MC}(\rho)$ is the Monte Carlo EEDF.}
\label{larger-fit1}
\end{table}

\begin{table}
\footnotesize
\begin{center}
\begin{tabular}{|c|c|c||c|c|c|}
\hline
$N$ & $N_{sh}$ & &$\rho_{\rm min} = 1.$ & $\rho_{\rm min} = 1.5 $& $\rho_{\rm min} = 2.$\\
\hline
500 & 1 & $\sigma$ &
   0.25638(80) &   0.2468(11) &   0.2237(17) \\
& & $f_\infty$ &
   0.015976(11) &   0.016130(17) &   0.016540(29) \\
& 10 & $\sigma$ &
   0.25607(81) &   0.2463(11) &   0.2230(17) \\
& & $f_\infty$ &
   0.015976(11) &   0.016132(17) &   0.016548(29) \\
\hline
1000 & 1 & $\sigma$ &
   0.25146(64) &   0.24592(89) &   0.2304(14) \\
& & $f_\infty$ &
   0.016019(9) &   0.016108(13) &   0.016382(23) \\
& 20 & $\sigma$ &
   0.25030(65) &   0.24418(90) &   0.2276(14) \\
& & $f_\infty$ &
   0.016027(9) &   0.016126(14) &   0.016420(23) \\
\hline
2000 & 40 & $\sigma$ &
   0.24691(61) &   0.24285(85) &   0.2324(13) \\
& & $f_\infty$ &
   0.016062(9) &   0.016128(13) &   0.016311(22) \\
& 80 & $\sigma$ &
   0.24688(63) &   0.24276(87) &   0.2325(13) \\
& & $f_\infty$ &
   0.016062(9) &   0.016129(13) &   0.016310(22) \\
\hline
4000 & 100 & $\sigma$ &
   0.24530(59) &   0.24121(82) &   0.2351(13) \\
& & $f_\infty$ &
   0.016074(8) &   0.016141(12) &   0.016247(21) \\
& 200 & $\sigma$ &
   0.24532(60) &   0.24115(84) &   0.2350(13) \\
& & $f_\infty$ &
   0.016073(8) &   0.016141(13) &   0.016248(22) \\
\hline
8000 & 200 & $\sigma$ &
   0.24274(50) &   0.24240(69) &   0.2376(11) \\
& & $f_\infty$ &
   0.016103(7) &   0.016109(10) &   0.016194(18) \\
& 800 & $\sigma$ &
   0.24278(53) &   0.24238(73) &   0.2375(11) \\
& & $f_\infty$ &
   0.016102(7) &   0.016109(11) &   0.016194(19) \\
\hline
16000 & 800 & $\sigma$ &
   0.24322(75) &   0.2434(11) &   0.2419(16) \\
& & $f_\infty$ &
   0.016092(11) &   0.016088(16) &   0.016116(27) \\
& 1200 & $\sigma$ &
   0.24330(77) &   0.2434(11) &   0.2415(16) \\
& & $f_\infty$ &
   0.016090(11) &   0.016088(16) &   0.016122(27) \\
\hline
32000 & 5000 & $\sigma$ &
   0.2413(13) &   0.2412(18) &   0.2440(28) \\
& & $f_\infty$ &
   0.016115(18) &   0.016115(27) &   0.016066(47) \\
\hline
\end{tabular}
\caption{Results for the fit  $f_{MC}(\rho) \exp(D \rho^\delta)
 = f_{\infty} \rho^\sigma$
for $\rho > \rho_{\rm min}$. $f_{MC}(\rho)$ is the Monte Carlo EEDF.
$D$ and $\delta$ have been set equal to the theoretical predictions,
Eqs. (\protect\ref{delta}) and (\protect\ref{Dconstant}).
Data with $\rho > 4.5$ have not been included in the fit.}
\label{larger-fit2}
\end{center}
\end{table}

\begin{table}
\footnotesize
\begin{center}
\begin{tabular}{|c|c|c||c|c|c|c|}
\hline
$N$ & $N_{sh}$ & & $\rho_{\rm max} = 0.8$ & $\rho_{\rm max} = 0.6 $& $\rho_{\rm max} = 0.4$& $\rho_{\rm max} = 0.2$\\
\hline
\hline
500 & 1 & $\theta$ &
    0.2177(53) &    0.2680(87) &    0.303(19) &    0.50(13) \\
& &$ f_1$ &
    0.01470(5) &    0.01542(11) &    0.01610(38) &    0.0228(54) \\
\hline
1000 & 1 &$ \theta$ &
    0.1975(41) &    0.2453(67) &    0.274(14) &    0.339(55) \\
& &$ f_1$ &
    0.01474(4) &    0.01543(9) &    0.01600(27) &    0.0179(18) \\
\hline
2000 & 1 &$ \theta$ &
    0.1906(39) &    0.2315(63) &    0.270(13) &    0.194(73) \\
& &$ f_1$ &
    0.01479(3) &    0.01539(8) &    0.01615(27) &    0.0140(19) \\
\hline
4000 & 1 &$ \theta$ &
    0.1895(37) &    0.2258(60) &    0.284(13) &    0.343(61) \\
& &$ f_1$ &
    0.01479(3) &    0.01536(8) &    0.01654(26) &    0.0185(21) \\
\hline
8000 & 1 &$ \theta$ &
    0.1874(31) &    0.2275(51) &    0.251(11) &    0.313(53) \\
& &$ f_1$ &
    0.01489(3) &    0.01546(7) &    0.01593(21) &    0.0177(17) \\
\hline
16000 & 1 &$ \theta$ &
    0.1721(46) &    0.2000(74) &    0.186(15) &    0.182(75) \\
& &$ f_1$ &
    0.01487(4) &    0.01527(10) &    0.01497(29) &    0.0151(21) \\
& 10 &$ \theta$ &
    0.1745(46) &    0.2035(75) &    0.191(16) &    0.189(76) \\
& &$ f_1$ &
    0.01486(4) &    0.01528(10) &    0.01500(29) &    0.0152(21) \\
& 20 &$ \theta$ &
    0.1750(46) &    0.2045(75) &    0.194(15) &    0.199(77) \\
& &$ f_1$ &
    0.01486(4) &    0.01529(10) &    0.01506(29) &    0.0154(22) \\
\hline
32000 & 1 &$ \theta$ &
    0.1416(75) &    0.178(12) &    0.212(26) &    0.28(13) \\
& &$ f_1$ &
    0.01482(7) &    0.01535(16) &    0.01609(51) &    0.0179(42) \\
& 10 &$ \theta$ &
    0.1579(75) &    0.200(12) &    0.243(26) &    0.36(13) \\
& &$ f_1$ &
    0.01480(7) &    0.01542(16) &    0.01637(52) &    0.0198(47) \\
& 20 &$ \theta$ &
    0.1597(75) &    0.203(12) &    0.247(26) &    0.38(13) \\
& &$ f_1$ &
    0.01480(7) &    0.01543(16) &    0.01640(52) &    0.0203(48) \\
\hline
\end{tabular}
\end{center}
\caption{Results for the fit  $f_{MC}(\rho)
 = f_1 \rho^\theta$
for $\rho < \rho_{\rm max}$. $f_{MC}(\rho)$ is the Monte Carlo EEDF.}
\label{smallr}
\end{table}


\begin{thebibliography}{199}
\bibitem{footnote0} 
The different proposals appearing in the literature and many 
old numerical results have been reviewed by 
Bishop and Clarke \cite{Bishop-Clarke_91}.
\bibitem{footnote4} The coupling constant $g$ can be expressed in 
terms of the second virial coefficient for a polydisperse ensemble
of polymers, see \cite{Nickel-Muthukumar}.
\bibitem{footnote5}
The most precise Monte Carlo estimate of $\Delta$ is
$\Delta = 0.517\pm 0.007 + 0.010$ \cite{Belohorec-Nickel_97}.
Perturbative field theory gives a lower estimate
\cite{Guida-ZinnJustin_98}: 
$\Delta = 0.478\pm 0.010$ ($d=3$ expansion) and 
$\Delta = 0.486 \pm 0.016$ ($\epsilon$-expansion).
\end{thebibliography}
\end{document}